\documentclass[aps,twocolumn,10pt,superscriptaddress,prb]{revtex4-1}

\usepackage{amsmath,amssymb,amsfonts}
\usepackage{graphicx}
\usepackage{braket}
\usepackage{nicefrac}
\usepackage{multirow}
\usepackage[thinlines]{easytable}
\usepackage{cellspace}
\usepackage[colorlinks=True,linkcolor=red,citecolor=blue,urlcolor=blue]{hyperref}
\usepackage[retainorgcmds]{IEEEtrantools}
\usepackage{xcolor}
\usepackage[title]{appendix}
\usepackage{soul}
\usepackage{url}

\begin{document}

%%%%%%%%%%%%%%%%%%%%%%%%%%%%%%%%%%%%%%%%%%%%%%%%%
% Title options
%%%%%%%%%%%%%%%%%%%%%%%%%%%%%%%%%%%%%%%%%%%%%%%%%
% - Analytic computation of SU(3) matrix elements
% - Removing the phase ambiguity of
%   quantum eigenstates
% - Removing the gauge arbitrariness of
%   quantum eigenstates

\title{Anatomy of $\mathbb{Z}_{2}$ fluxes in anyon Fermi liquids and Bose condensates}
%\title{Flux deconfinement in $\mathbb{Z}_{2}$ anyon Fermi liquids and anyon Bose condensates}

%\author{Oscar Pozo}
%\affiliation{Instituto de Ciencia de Materiales de Madrid, and CSIC, Cantoblanco, 28049 Madrid, Spain}
\author{Oscar Pozo}
\affiliation{Instituto de Ciencia de Materiales de Madrid,
and CSIC, Cantoblanco, 28049 Madrid, Spain}
\affiliation{Max-Planck Institute for the Physics of Complex Systems, D-01187 Dresden, Germany}

\author{Peng Rao}
\affiliation{Max-Planck Institute for the Physics of Complex Systems, D-01187 Dresden, Germany}

\author{Chuan Chen}
\affiliation{Max-Planck Institute for the Physics of Complex Systems, D-01187 Dresden, Germany}

\author{Inti Sodemann}
\affiliation{Max-Planck Institute for the Physics of Complex Systems, D-01187 Dresden, Germany}

\begin{abstract}
    We study in detail the properties of $\pi$-fluxes embedded in a state with a finite density of anyons that form either a Fermi liquid or a Bose-Einstein condensate. By employing a recently developed exact lattice bosonization in 2D, we demonstrate that such $\pi$-flux remains a fully deconfined quasiparticle with a finite energy cost in a Fermi liquid of emergent fermions coupled to a $\mathbb{Z}_2$ gauge field. This $\pi$-flux is accompanied by a screening cloud of fermions, which in the case of a Fermi gas with a parabolic dispersion binds exactly $1/8$ of a fermionic hole. In addition there is a long-ranged power-law oscillatory disturbance of the liquid surrounding the $\pi$-flux akin to Friedel oscillations. These results carry over directly to the $\pi$-flux excitations in orthogonal metals. In sharp contrast, when the $\pi$-flux is surrounded by a Bose-Einstein condensate of particles coupled to a $\mathbb{Z}_{2}$ gauge field, it binds a superfluid half-vortex, becoming a marginally confined excitation with a logarithmic energy cost divergence.
\end{abstract}

\maketitle

\section{Introduction}

$\mathbb{Z}_{2}$ topologically ordered states are one of the most well understood fractionalized states of matter~\cite{wen2004quantum,fradkin2013field}. They were first introduced by Anderson, in the form of short-ranged resonant-valence-bond (RVB) spin-liquid states~\cite{anderson1987resonating,baskaran1987resonating}, and shortly after it was understood~\cite{read1989statistics,kivelson1989statistics,read1991large,sachdev1991large} that they possess non-local quasiparticles, one being the spinon and the other a $\pi$-flux tube for the spinon, called the vison~\cite{senthil2000z}. These early works found out that the statistics of the spinon could be transmuted from fermionic to bosonic depending on whether the $\pi$-flux was attached to it or not~\cite{read1989statistics,kivelson1989statistics}. The utility of viewing the $\pi$-flux as a bosonic quasiparticle that can be condensed to transit from the $\mathbb{Z}_{2}$ topologically ordered states towards a variety of conventionally ordered states was later developed and exploited in a series of works~\cite{balents1998nodal,balents1999dual,senthil2000z}.

A remarkably simple incarnation of $\mathbb{Z}_{2}$ topological order was introduced by Kitaev in his Toric Code (TC) model Hamiltonian~\cite{kitaev2003fault}, whose ground states and its entire excitation spectrum can be solved for exactly. Because the TC has no global SU(2) symmetry, the particle that is viewed as the spinon or the $\pi$-flux is a matter of convention. These two particles are often denoted by $e$ (electric boson) and $m$ (magnetic boson). Their bound or fused state is fermionic and it is denoted by $\varepsilon$. By elaborating on observations from Ref.~[\onlinecite{gaiotto2016spin}], an interesting perspective of the TC has been developed recently in Ref.~[\onlinecite{chen2018exact}], which exploits the TC quasiparticle structure to provide a precise 2D lattice implementation of bosonization preserving locality. For related ideas and elaborations see also Refs.~[\onlinecite{bravyi2002fermionic,levin2003fermions,verstraete2005mapping,ball2005fermions,levin2006quantum,chen2019bosonization,radicevic2018spin,chen2019exact,RaoInti2020}]. The idea is basically an extension of the 1D Jordan-Wigner transformation to 2D, in which ordinary fermionic models can be mapped onto 2D spin TC-like models, by viewing the fermions as the $\varepsilon$ particles added on top of the TC vacuum. Importantly, this mapping preserves the spatial locality of the Hamiltonian, in contrast with other 2D lattice bosonization approaches, such as some lattice implementations of Chern-Simons theories~\cite{fradkin1989jordan,eliezer1992intersection,eliezer1992anyonization,lopez1994chern,kumar2014chern}, which map a local fermionic model into a bosonic model with non-local interactions or viceversa. 
For other 2D lattice bosonization constructions preserving locality see Refs.~[\onlinecite{chen2007quantum,feng2007topological,chen2007exact,chen2008exact,chen2019bosonization,chen2019exact}].

\begin{figure}[!t]
    \centering
    \includegraphics[scale=1.0]{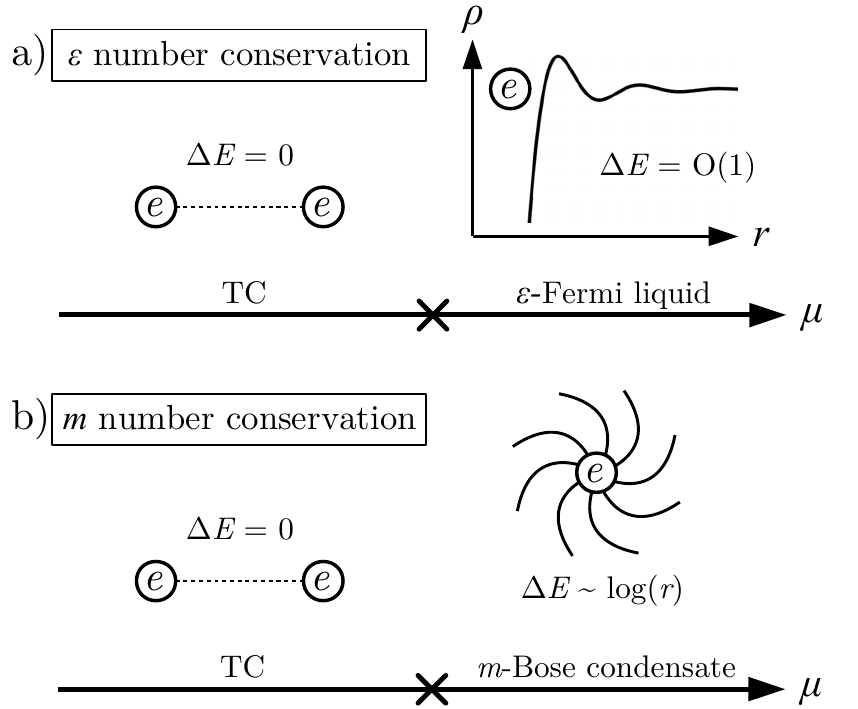}
    \caption{By enriching the Toric Code with particle number conservation of one of its anyons, we can construct a) Fermi liquids of $\varepsilon$ particles where the $e$ particle remains fully deconfined, or b) Bose condensates of $m$ particles where the $e$ particle is logarithmically confined by binding a half-vortex.}
    \label{fig:Fig-0}
\end{figure}

The key ingredient in the bosonization approach of Ref.~[\onlinecite{chen2018exact}] is the enforcement of a local conservation law (gauge symmetry) in the TC Hamiltonian that freezes the motion of isolated $e$ and $m$ particles which would otherwise have long-ranged statistical interactions with the $\varepsilon$ particle. As a result, the Hilbert space reduces to a direct sum of subspaces in which the $\varepsilon$ particle is allowed to fluctuate but its dynamics can be described as if it was an ordinary local fermion. This ingredient is essentially the same that allows for the exact solvability of the Kitaev honeycomb model~\cite{kitaev2006anyons}, which can be viewed as a special case of the models in Ref.~[\onlinecite{chen2018exact}]. This construction can also be considered as the fermionic version of the exact Kramers-Wannier-type duality of conventional $\mathbb{Z}_{2}$ lattice gauge theory~\cite{wegner1971duality,fradkin1978order}, where one enforces a local symmetry that fixes the vacuum to not contain $\mathbb{Z}_{2}$ bosonic charges and only allows to pair-fluctuate the bosonic $\pi$-fluxes~\cite{fradkin2013field,wen2004quantum}, as we will review in Sec.~\ref{sec:Sec2B}.

% Is it clear that the $m$ particle sees $e$ particles as pi-fluxes and viceversa?
In this work, we exploit and generalize these ideas by taking the ground state of the TC model as a convenient vacuum to construct phases of matter with a finite density of its anyonic particles (see Fig.~\ref{fig:Fig-0}). We perform these constructions in a microscopically explicit form by endowing the anyons with a global U(1) symmetry. Let us now summarize the contents of our manuscript and the main results. In an effort to make our presentation self-contained, we begin in Sec.~\ref{sec:ToricCodeZ2} by reviewing basic aspects of the TC. In Sec.~\ref{sec:BoseZ2}, we consider the bosonic case in which one of the bosonic anyons, for instance the $m$ particle, undergoes a chemical potential driven phase transition into a superfluid phase, while the $e$ particle remains gapped and dynamically frozen, acting as a $\pi$-flux source. We demonstrate that an added isolated $e$ particle binds a half-integer vortex of the surrounding $m$-particle superfluid, therefore implying that it is marginally confined with a logarithmically divergent energy cost. This constitutes a mechanism for vortex fractionalization in a superfluid that nonminimally has a Bose condensate of charge 1, as originally pointed out by Kivelson in Ref.~[\onlinecite{kivelson1989statistics}]. See also Ref.~[\onlinecite{kitazawa1990topological}] for a closely related discussion.

In Sec.~\ref{sec:TC-FermionicZ2} we consider the case of a Fermi liquid of $\varepsilon$ particles. To construct such Fermi liquid states we endow the $\varepsilon$ particles with a global U(1) particle number conservation symmetry. The resulting Fermi liquid of $\varepsilon$ particles therefore shares many universal properties with orthogonal metals~\cite{nandkishore2012orthogonal}, and some of our results carry over to these states. However, these phases of matter are not strictly the same, because in our case the U(1) symmetry is not the microscopic electron number conservation. In Section~\ref{sec:Z2FermiLiquids} we study the properties of the single static $e$ particle embedded in such Fermi liqudis of $\varepsilon$ particles, where it acts as a $\pi$-flux source, and we find that they display various remarkable properties. We find that this $\pi$-flux remains as a fully deconfined finite energy excitation in the presence of the surrounding fermionic fluid. Additionally, we show that the flux acquires a characteristic 
``screening'' cloud of $\varepsilon$ fermions. When the fermions have a parabolic dispersion, this cloud contains a depletion of exactly $1/8$ of $\varepsilon$ fermions. The $\pi$-flux also induces static spatial oscillations of the density of $\varepsilon$ fermions with wavevector $2k_{F}$ and that decay as a power law $\sim 1/r^{2}$ in analogy to Friedel oscillations~\cite{giuliani2005quantum}. We also study the properties of this $\pi$-flux on the Fermi liquid in the square lattice for all fillings, demonstrating that it always remains a deconfined quasiparticle and computing the explicit dependence of the number of $\varepsilon$ fermions that form the screening cloud surrounding the $\pi$-flux.  This $\pi$-flux deconfinement carries over directly to the case of orthogonal metals, where the fraction of fermions making up the screening cloud translates into an amount of physical electric charge surrounding the vison excitations.
%in such states. 
Finally, we close in Section~\ref{sec:Discussion} with a summary and discussion of our results.

\section{Toric Code and bosonic $\mathbb{Z}_{2}$ Lattice Gauge Theory}
\label{sec:ToricCodeZ2}

\subsection{Toric Code review}
\begin{figure}[!]
    \centering
    \includegraphics{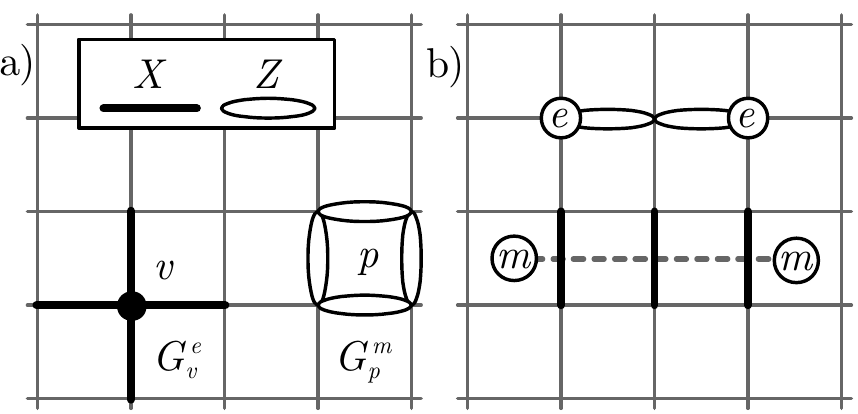}
    \includegraphics{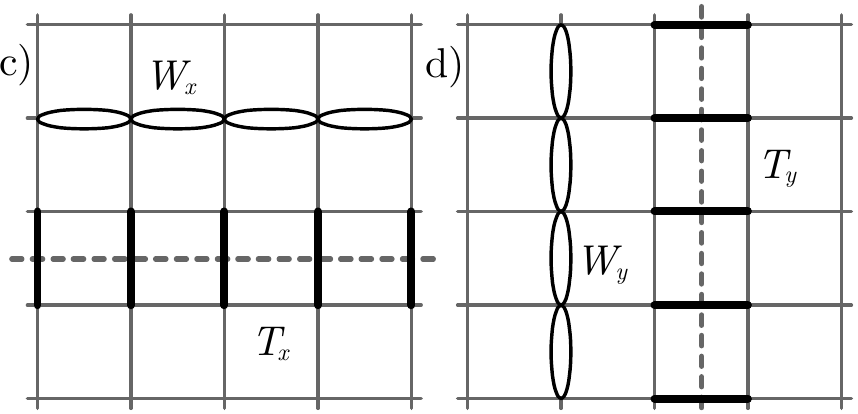}
    \caption{The Toric Code model, with spin-1/2 degrees of freedom residing in the links of a square lattice. a) Vertex and plaquette operators $G_{v}^{e},G_{p}^{m}$ defined in Eq.~\eqref{eq:BasicLocalOp}. 
    %The links where an $X$ operator is applied are highlighted in black, while $Z$ operators are indicated with an empty ellipsoid. 
    b) Pairs of $e$ ($m$) particles are created by applying strings of $Z_{l}$ ($X_{l}$) operators to the ground state. c),d) Non-contractible t'Hooft and Wilson loop operators defined in Eq.~\eqref{eq:NCLoopOp} in the $\hat{x}$ and $\hat{y}$ directions, respectively.}
    \label{fig:Fig1}
\end{figure}

% Introduce the TC
We will exploit the Toric Code~\cite{kitaev2003fault} as a vacuum to construct states. Its Hamiltonian acts on spin-1/2 degrees of freedom residing on the links of a two-dimensional (2D) square lattice placed on a torus and it is given by
\begin{IEEEeqnarray}{rCl}
    H & = & \Delta_{m}\sum_{\mathrm{all} \ p}\left( \dfrac{1-G_{p}^{m}}{2} \right) + \Delta_{e}\sum_{\mathrm{all} \ v}\left( \dfrac{1-G_{v}^{e}}{2} \right) \ , \label{eq:TCHam} \\
    G_{v}^{e} & = & \prod_{l\in v} X_{l}  \ , \qquad
    G_{p}^{m} = \prod_{l\in p} Z_{l}  \ , 
    \label{eq:BasicLocalOp}
\end{IEEEeqnarray}
where $X,Y,Z$ denote Pauli matrices, $\Delta_{m,e}>0$ and subindices $l,v,p$ label the links, vertices and plaquettes of the lattice, respectively. The links involved in the operators $G_{v}^{e},G_{p}^{m}$ are illustrated in Fig.~\ref{fig:Fig1}a). These local operators commute and have eigenvalues $\pm1$. In the ground state all $G_{v}^{e},G_{p}^{m}$ take the value $+1$, and any local excitation has an energy gap $\Delta_{e,m}$. We say that an $e$ ($m$) particle resides in a vertex $v$ (plaquette $p$) if $G_{v}^{e}=-1$ ($G_{p}^{m}=-1$) and otherwise we say it is empty. The application of $X_{l}$ and $Z_{l}$ operators to the ground state creates pairs of these particles in the adjacent vertices and plaquettes, as shown in Fig.~\ref{fig:Fig1}b). There are two global constraints on these operators on a 2D torus
\begin{equation}
    \prod_{\mathrm{all} \ v} G_{v}^{e} = 1 \ , \qquad \prod_{\mathrm{all} \ p} G_{p}^{m} = 1 \ ,
    \label{eq:ParityConstraint}
\end{equation}
which imply that the number of $e$ or $m$ particles must be even in any state. In addition, we can construct the following four $\mathbb{Z}_{2}$ loop operators that commute with the Hamiltonian. They are depicted in Fig.~\ref{fig:Fig1}c),d) and given by
\begin{equation}
    T_{x,y} = \prod_{ \ l \ \in \ \beta_{x,y}} X_{l} \ , \qquad W_{x,y} = \prod_{ \ l \ \in \ \gamma_{x,y}} Z_{l} \ ,
    \label{eq:NCLoopOp}
\end{equation}
where the links belonging to the non-contractible loops $\beta_{x,y}$ and $\gamma_{x,y}$ along the $x,y$ directions of the torus are shown in Fig.~\ref{fig:Fig1}c),d). These loop operators are also known as t'Hooft and Wilson loops, and importantly, they cannot be written in terms of the local operators $G_{v}^{e},G_{p}^{m}$. The specific shape of the loop defining $T_{x,y}$ and $W_{x,y}$ can be altered by multiplying the operators defined in Eq.~\eqref{eq:NCLoopOp} by $G_{v}^{e}$ and $G_{p}^{m}$ operators, respectively.
%are global operators that cannot be written in terms of local $G_{v}^{e},G_{p}^{m}$ operators. Their specific form can be varied by locally applying $G_{v}^{e},G_{p}^{m}$ operators, 
%so the only physical content of these global operators is found on their non-contractible direction. 
%so only their non-contractible loop direction has a physical meaning.
They satisfy the following algebra
\begin{IEEEeqnarray}{rClrClrCl}
    \left\{ T_{x},W_{y} \right\} & = & 0 \ , & \quad \left[ T_{x},W_{x} \right] & = & 0 \ , & \quad \left[ T_{x},T_{y} \right] & = & 0 \ , \quad \\
    \left\{ T_{y},W_{x} \right\} & = & 0 \ , & \quad \left[ T_{y},W_{y} \right] & = & 0 \ , & \quad \left[ W_{x},W_{y} \right] & = & 0 \ , \quad
\end{IEEEeqnarray}
which implies that all states, including the ground state, are 4-fold degenerate. As depicted in Fig.~\ref{fig:Fig1}, these global operators can be interpreted as the operators that create a pair of $e$ or $m$ particles from the vacuum and split them over a non-contractible loop of the torus to finally annihilate them.

%\textst{Although these loop operators are independent of $G_{v}^{e},G_{p}^{m}$, their commutativity with the Hamiltonian follows from the commutativity of every $G_{v}^{e}, G_{p}^{m}$ with the Hamiltonian, respectively}.\\

\subsection{Bosonic $\mathbb{Z}_{2}$ lattice gauge theory}
\label{sec:Sec2B}
% How do we get a $\mathbb{Z}_{2}$ gauge theory from Toric Code

%1 - General, and without $e$ particles

%\noteRP{Note that $G_{p}^{m}$ does not commute with the Hamiltonian, $m$ particles can move and their occupation density is not fixed. The remaining degrees of freedom are not specified by $G_{p}^{m}=\pm1$.}

The Hamiltonian from Eq.~\ref{eq:TCHam} describes $e$ and $m$ particles as gapped excitations with no dynamics. To view the TC as a $\mathbb{Z}_{2}$ lattice gauge theory
%~\cite{wegner1971duality} 
we reduce its symmetries by allowing dynamics just for $m$ particles, while keeping $e$ particles gapped and frozen. 
This is realized by imposing a local conservation law for the operator that measures the presence of $e$ particles at each vertex, namely $\left[ H, G_{v}^{e} \right] = 0$ for any $v$. 
Although the dynamics of $m$ particles spoils the commutation relation $\left[ H,W_{x,y} \right]$ making it non-zero, the loop operators $T_{x,y}$ still commute with the Hamiltonian. Therefore, this gauge structure divides the Hilbert space into subspaces in which the eigenvalues of $\left\{ G_{v}^{e},T_{x},T_{y} \right\}$ are fixed. The remaining degrees of freedom correspond to the $m$ particle dynamics, which is additionally subject to the global particle parity constraint from Eq.~\eqref{eq:ParityConstraint}. These degrees of freedom can be encoded in a dual representation in which we assign a spin-1/2 variable, or equivalently a hardcore boson $n_{p} = b_{p}^{\dagger}b_{p}\in\{ 0,1 \}$ to each plaquette. The four-spin operator $G_{p}^{m}$ from Eq.~\eqref{eq:BasicLocalOp} maps to the dual spin operator $Z_{p}$, measuring the parity of the $m$ particle in a plaquette as follows:
\begin{IEEEeqnarray}{rCl}
    G_{p}^{m} \Longleftrightarrow Z_{p} & = & (-1)^{n_{p}} \ ,
    \label{eq:Gpdual}
\end{IEEEeqnarray}
In addition, the constraint on the global parity of $m$ particles from Eq.~\eqref{eq:ParityConstraint} is represented as a dual global $\mathbb{Z}_{2}$ parity ``symmetry'':
\begin{equation}
    \prod_{\mathrm{all} \ p} (-1)^{n_{p}} = 1 \ .
    \label{eq:Z2ParitySymmetry}
\end{equation}
This implies that, in this dual picture, parity-even states have a one-to-one correspondence with physical states while parity-odd states must be discarded as unphysical. The dual representation of the pair creation operator of $m$ particles depends on which Hilbert subspace $\left\{ G_{v}^{e},T_{x},T_{y} \right\}$ we are considering. In the simplest case there are no $e$ particles and the loop operators have trivial eigenvalues, namely $\left\{ G_{v}^{e},T_{x},T_{y} \right\} = \left\{ 1,1,1 \right\}$. In this case, the pair creation operator of $m$ particles can be represented as
\begin{equation}
    X_{l} \Longleftrightarrow X_{p(l)}X_{p'(l)} = \left( b_{p(l)}+b_{p(l)}^{\dagger} \right) \left( b_{p'(l)}+b_{p'(l)}^{\dagger} \right) \ .
    \label{eq:Xdual}
\end{equation}
Here and in the following, the plaquettes adjacent to the link $l$ are labeled as $p(l)$ and $p'(l)$. The set of operators $\left\{ G_{p}^{m},X_{l} \right\}$ provides a complete basis to algebraically construct all the operators that commute with $G_{v}^{e}$. Notice that Eq.~\eqref{eq:Xdual} automatically implies that $\left\{ G_{v}^{e},T_{x},T_{y} \right\} = \left\{ 1,1,1 \right\}$.

% 2 - Include $e$ particles

\begin{figure}[!]
    \centering
    \includegraphics{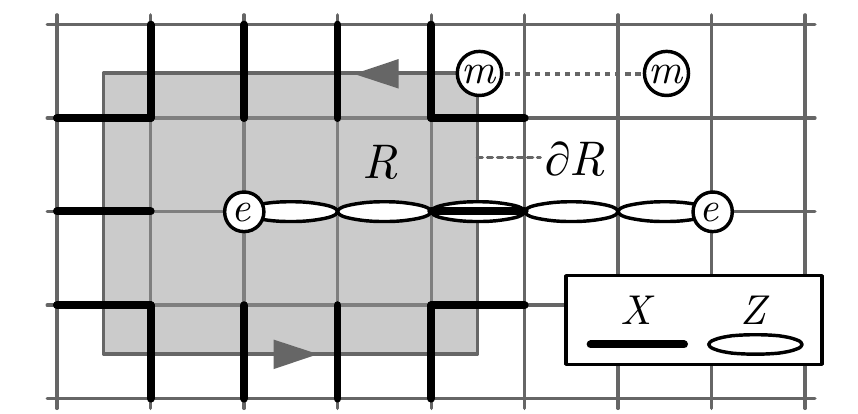}
    \caption{The braiding of an $m$ particle along the boundary of an area $R$ results in a global minus sign for the final state that equals the action of the operator $\Pi_{e}$ over $R$, which measures the $e$ particle number parity inside $R$, as defined in Eq.~\eqref{eq:eAreaParity}. In the dual representation there is a branch cut in the hopping of $m$ particles across the white bonds that connect the two $e$ particles.}
    \label{fig:Fig2}
\end{figure}

The dual representation from Eq.~\eqref{eq:Xdual} must be modified for the Hilbert subspaces containing $e$ particles. To do this, we begin by noticing that the operator that measures the parity of $e$ particles inside a region $R$ is given by
\begin{equation}
    \Pi_{e} \equiv \prod_{v \in R}G_{v}^{e} \ .
    \label{eq:eAreaParity}
\end{equation}
When $R$ is a simply connected region, this operator is equivalent to a closed loop operator that transports the $m$ particle along the boundary of $R$, as depicted in Fig.~\ref{fig:Fig2}. However, notice that Eq.~\eqref{eq:Xdual} would imply that the transport of $m$ in the dual representation is always equal to $1$. To correct for this, the representation of the transport operator of $m$ particles must encode the statistical interaction with the $e$ particle. This can be accomplished by drawing lines that connect each pair of $e$ particles as depicted in Fig.~\ref{fig:Fig2}.
These lines are ``branch cuts'' in which we flip the sign of the representation in Eq.~\eqref{eq:Xdual} as follows
\begin{equation}
    X_{l} \Longleftrightarrow e^{iA_{l}}X_{p(l)}X_{p'(l)} \ ,
    \label{eq:DualityRelationX}
\end{equation}
where $A_{l}=\pi$ when the link $l$ belongs to the branch cut and $0$ otherwise (see Fig.~\ref{fig:Fig2}). In other words, the $m$ particles see the $e$ particles as localized sources of $\pi$-flux.

% 3 - Include topology

Similarly, when $T_{x,y}=-1$, we introduce antiperiodic boundary conditions for the dual bosons along the corresponding direction in the torus. The precise choice for the dual vector potential associated with the $e$ particles or the aforementioned twists of the boundary conditions around the torus is a matter of convention that can be modified provided its integrals along closed loops are unchanged modulo $2\pi$. These refinements complete an exact dual hardcore boson representation, summarized in Eqs.~\eqref{eq:Gpdual} and \eqref{eq:DualityRelationX}, for the full Hilbert space of any microscopic spin Hamiltonian that commutes with $G_{v}^{e}$ at every vertex.

\subsection{Spontaneous symmetry breaking of the unphysical $\mathbb{Z}_{2}$ symmetry}
\label{sec:Sec2C}

\begin{figure*}[!t]
    \centering
    \includegraphics[scale=1.0]{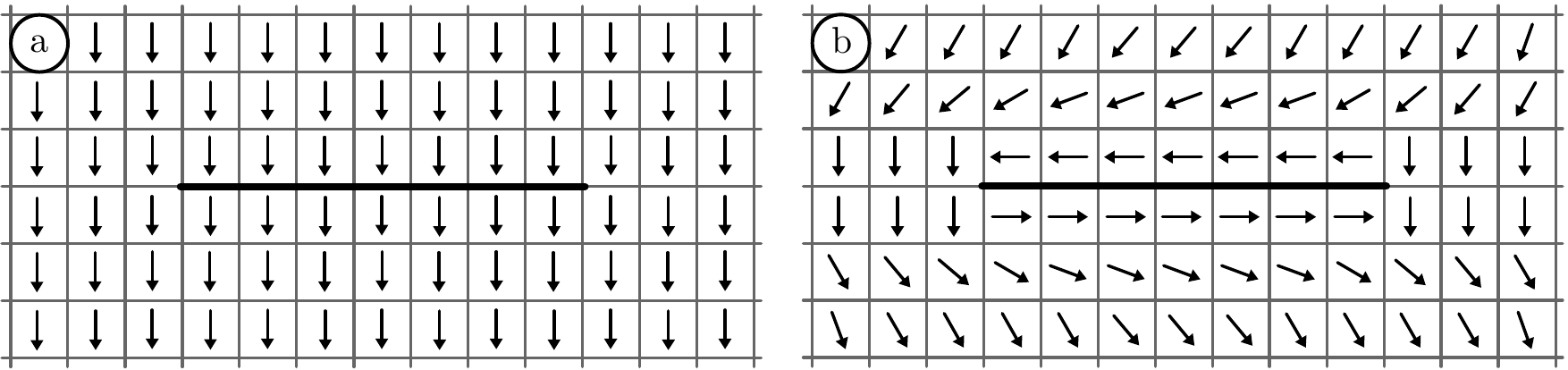}
    \caption{Pair of $e$ particles (ends of solid black string) in a) dual Ising magnet and b) superfluid or XY ferromagnet of $m$ particles. Arrows indicate the orientation of the dual XY order parameter $\braket{X+iY}$ (superfluid phase). The branch cut that ties the $e$ pair forces neighboring phases on the surrounding fluid to be anti-aligned. The Ising magnet cannot smoothly adapt to it and carries an energy cost proportional to the branch cut length. On the contrary, the superfluid adapts to it reducing the energy cost from linear to logarithmic, and recovers a global alignment of the phases far away from the $e$ pair.}
    \label{fig:Fig4}
\end{figure*}

\begin{figure}[!t]
    \centering
    \includegraphics{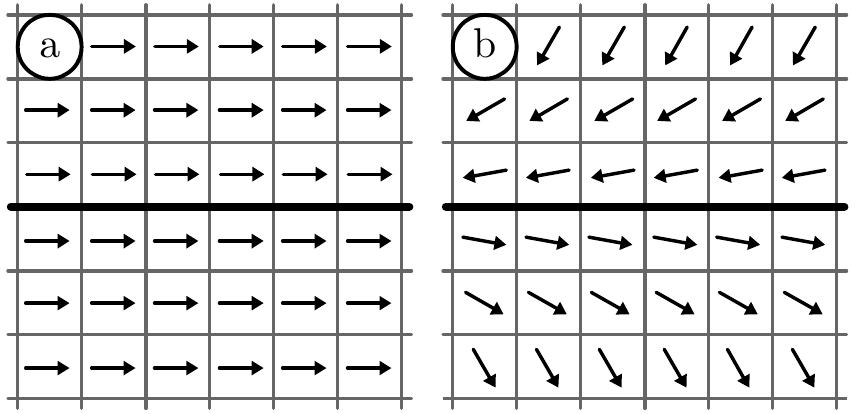}
    \caption{Twisted boundary conditions in a torus for a) a dual Ising magnet and b) a superfluid or XY ferromagnet. Vectors depict the boson wave function phase at each plaquette. The finite plane is viewed as a torus in both directions by identifying links in the edges. The branch cut, highlighted in black, forces neighboring phases to be anti-aligned. The Ising magnet cannot smoothly adapt to such branch cut and carries an energy cost proportional to the branch cut length. On the contrary, the superfluid adapts to such a twist leading to a constant finite energy cost in the thermodynamic limit (see Eqs.~\eqref{eq:EnTwist1}-\eqref{eq:EnTwist3}).}
    \label{fig:Fig3}
\end{figure}

Remarkably, in spite of the parity ``symmetry'' of $m$ particles from Eq.~\eqref{eq:Z2ParitySymmetry} not being a physical symmetry but rather a constraint, this ``symmetry'' can be spontaneously broken because it is only a global one, and therefore escapes the constraints of Elitzur's theorem~\cite{elitzur1975impossibility}. The classic Hamiltonian to illustrate this phenomenon is
\begin{IEEEeqnarray}{rCl}
    H & = & \Delta_{m}\sum_{\mathrm{all} \ p}\left( \dfrac{1-G_{p}^{m}}{2} \right) + \Delta_{e}\sum_{\mathrm{all} \ v}\left( \dfrac{1-G_{v}^{e}}{2} \right) - \nonumber \\
    & & -t\sum_{\mathrm{all} \ l} X_{l} \ ,
    \label{eq:HamDirectPhysicalPicture}
\end{IEEEeqnarray}
where $t\geq0$. Using the previous dual representation from Eqs.~\eqref{eq:Gpdual} and \eqref{eq:DualityRelationX} in the absence of $e$ particles and for periodic boundary conditions, $\left\{ G_{v}^{e},T_{x},T_{y} \right\}= \left\{ 1,1,1 \right\} $, this Hamiltonian maps to the familiar transverse field Ising model
\begin{equation}
    H = -t\sum_{\langle p,p' \rangle}X_{p}X_{p'} - \dfrac{\Delta_{m}}{2}\sum_{\mathrm{all} \ p}Z_{p} \ ,
    \label{eq:DualHamWithoute}
\end{equation}
where $\langle p,p' \rangle$ denotes nearest-neighbor plaquettes and we have omitted a global additive constant. When $\Delta_{m}\ll t$ it is clear that this model spontaneously breaks the parity ``symmetry'' from Eq.~\eqref{eq:Gpdual}, by having an expectation value for the dual $X_{p}$ spin. In fact, in the limit of $\Delta_{m}=0$, there are two degenerate ground states with $X_{p}=\pm 1$. However, only the symmetric combination of these two states has $+1$ eigenvalue of the $\mathbb{Z}_{2}$ parity transformation from Eq.~\eqref{eq:Gpdual} and is therefore physical. In other words, one of the two states from the Anderson tower associated with the $\mathbb{Z}_{2}$ spontaneous symmetry breaking must be discarded as unphysical. Notice that the absence of degeneracy for $\Delta_{m}=0$ is completely transparent in the original physical picture from Eq.~\eqref{eq:HamDirectPhysicalPicture}, since in this case the ground state is a simple non-entangled direct product state with the spins at every link satisfying $X_{l}=1$. Now, when a pair of $e$ particles is placed in two vertices separated by a straight line $\gamma$, the dual Hamiltonian becomes
\begin{IEEEeqnarray}{rCl}
    H & = & -t\sum_{\langle p,p' \rangle \not\in \gamma} X_{p}X_{p'} + t\sum_{\langle p,p' \rangle \in \gamma} X_{p}X_{p'} + \nonumber \\
    & & + 2\Delta_{e} - \dfrac{\Delta_{m}}{2}\sum_{\mathrm{all} \ p} Z_{p} \ .
    \label{eq:14}
\end{IEEEeqnarray}
From Eq.~\eqref{eq:14} it directly follows that, for $\Delta_{m}=0$, the ground state still has all dual spins aligned in directions $X_{p}=\pm1$, but its energy is increased by $\Delta E = 2tL+2\Delta_{e}$ with respect to the ground state energy of Eq.~\eqref{eq:DualHamWithoute}, where $L$ is the integer measuring the length of the line $\gamma$ as depicted in Fig.~\ref{fig:Fig4}a). Therefore, there is a ``string tension'' that induces strong confinement between $e$ pairs in the dual broken symmetry state, in contrast to the Toric Code vacuum in which such pair of excitations are fully deconfined with a constant energy cost $\Delta E = 2\Delta_{e}$. A related signature of confinement appears when analyzing lowest energy states in sectors with twisted boundary conditions. Considering $(T_{x},T_{y})=(-1,1)$ for instance, its energy is $\Delta E = 2t L_{y}$ larger than the one with periodic boundary conditions as illustrated in Fig.~\ref{fig:Fig3}a). This is in sharp constrast to the TC Hamiltonian, where all the 4 sectors with twisted boundary conditions have the same energy, which is the characteristic topological degeneracy of the TC.
%\color{black}
We conclude that the ground state of our dual Hamiltonian has no deconfined $e$ particles and no topological ground state degeneracy in the torus when $t\ll\Delta_m$.

As we have reviewed, a spontaneous symmetry breaking transition of a discrete global unphysical symmetry, namely the parity of the $m$ particles, leads to a trivial phase in which all anyons experience strong linear confinement. Such transitions fall within the broader class of formal anyon ``bose-condensation'' transitions~\cite{kong2014anyon,burnell2018anyon}. The result of such formal ``condensation'' of a particle with self-bosonic statistics is a modification of the topological order of the system into a new one described by fusion rules in which the condensed particle is identified with the vacuum sector, $m\sim 1$, and also every other particle that braids non-trivially with such particle. In the current case, this leads therefore to a completely trivial state since all the other non-local anyons, $e$ and $\varepsilon$, braid non-trivially with $m$. However, as we will see in the next section, when the condensing boson carries a global U(1) quantum number such condensation does not lead to a completely trivial state, but instead one in which anyons that braid non-trivially with the condensing boson trap fractional vortices of the superfluid of the condensing particle, and thus remain marginally confined attracting each other with a potential that only increases logarithmically rather than linearly with distance.

\section{U(1) Symmetry Enriched $\mathbb{Z}_{2}$ Lattice Gauge Theory and Anyon Bose Condensation}
\label{sec:BoseZ2}

We now endow the structure of the bosonic $\mathbb{Z}_{2}$ lattice gauge theory with a \emph{global U(1) symmetry} that enforces the conservation of the total number of $m$ particles. We take the $m$ particle number to be
\begin{equation}
    n_{p} \equiv \dfrac{1-G_{p}^{m}}{2} \ , \qquad N_{m} = \sum_{\mathrm{all} \ p} n_{p} \ ,
\end{equation}
where $G_{p}^{m}$ is defined in Eq.~\eqref{eq:BasicLocalOp}. The structure of the Hilbert space is the same described in the previous section, except that now the subspaces defined by the number and position of $e$ particles and the twist of boundary conditions for $m$ particles, are further split into subspaces with definite total $m$ particle numbers specified by $N_{m}$. The parity symmetry from Eq.~\eqref{eq:ParityConstraint} is a subgroup of the U(1) symmetry generated by $N_{m}$. This parity subgroup still plays a special role, since every state with an odd number of $m$ particles must be discarded as unphysical. The allowed terms in the Hamiltonian must commute with the set $\left\{ N_{m}, G_{v}^{e} \right\}$, 
%\color{blue}
so they can only be the hopping and particle number operators. Any hopping operator between arbitrary plaquettes can be written in terms of nearest-neighbor hopping operators. Using the dual representation from Eqs.~\eqref{eq:Gpdual} and \eqref{eq:DualityRelationX}, we construct them as
%\color{black}
\begin{equation}
    e^{iA_{l}}b_{p(l)}^{\dagger}b_{p'(l)} \Longleftrightarrow \left( \dfrac{1-G_{p(l)}^{m}}{2} \right) X_{l} \left( \dfrac{1-G_{p'(l)}^{m}}{2} \right) . \quad
    \label{eq:HoppingDual}
\end{equation}
where again $p(l),p'(l)$ are the plaquettes adjacent to the link $l$, and the vector potential $A_{l}$ is chosen in the same way as discussed in Sec.~\ref{sec:Sec2B}. These hopping operators and $n_{p}$ form the basis for the algebra of local operators that preserves the $\mathbb{Z}_{2}$ gauge structure and the total number of $m$ particles. This structure is essentially a lattice version of the mutual Chern-Simons field theory that describes $\mathbb{Z}_{2}$ gauge theory~\cite{diamantini1996gauge,hansson2004superconductors,kou2008mutual}. More precisely, it only contains part of this structure since we have frozen $e$ particles, and thus their statistical influence on the hopping of $m$ particles is the only one captured in Eq.~\eqref{eq:HoppingDual} via the static gauge field $A_{l}$.

% SSB model, now with $N_{m}$ conserved.

% 1 - Dual model Hamiltonian

Again, because the $m$ particle number conservation is a \emph{global} symmetry it is perfectly legitimate to have phases in which it is spontaneously broken. A natural route to drive transitions into such phases is to proliferate the $m$ particle density via a chemical potential driven Bose-Einstein condensation. A paradigmatic model to realize this transition is the Bose-Hubbard model in the square lattice~\cite{fisher1989boson}
\begin{IEEEeqnarray}{rCl}
    H & = & -\dfrac{t}{2}\sum_{\mathrm{all} \ l} X_{l}\left( 1-G_{p(l)}^{m}G_{p'(l)}^{m} \right) + \nonumber \\ 
    & & + \Delta_{m}\sum_{\mathrm{all} \ p}\left( \dfrac{1-G_{p}^{m}}{2} \right) + \Delta_{e}\sum_{\mathrm{all} \ v}\left( \dfrac{1-G_{v}^{e}}{2} \right) . \qquad
\end{IEEEeqnarray}
In fact, in the trivial subspace with $\left\{ G_{v}^{e},T_{x},T_{y} \right\} = \left\{ 1, 1, 1 \right\}$, this Hamiltonian maps \emph{exactly} onto hardcore bosons on the square lattice or XXZ model 
%\noteOP{Factor 2 added to t in (18) and (19)}
\begin{IEEEeqnarray}{rCl}
    H & = & -t\sum_{\langle p,p' \rangle} \left( b_{p}^{\dagger}b_{p'} + b_{p'}^{\dagger}b_{p} \right) + \Delta_{m}\sum_{\mathrm{all} \ p} b_{p}^{\dagger}b_{p} \label{eq:XXZmodel1} \\
    & = & -\dfrac{t}{2}\sum_{\langle p,p' \rangle}\left( X_{p}X_{p'}+Y_{p}Y_{p'} \right) + \Delta_{m}\sum_{\mathrm{all} \ p}\left( \dfrac{1-Z_{p}}{2} \right) . \qquad
    \label{eq:XXZmodel2}
\end{IEEEeqnarray}
Increasing the hopping $t$ induces a chemical potential driven phase transition from the vacuum containing no $m$ particles to a state with a finite boson density, at $t_{c}=\Delta_{m}/4$, which we will also refer to as a boson proliferation transition. The ground state becomes a superfluid or XY ferromagnet of $m$ bosons with a finite stiffness since the bosons are always interacting due to their hardcore nature. This state breaks spontaneously the U(1) symmetry, including its $\mathbb{Z}_{2}$ parity subgroup. As before, only $\mathbb{Z}_{2}$ parity even states are physical.
% Here we assume for the phase transition that the chemical potential lies at $\mu=0$. Then when t=0 the energy bands are Toric Code like, and they are flat and separated $\Delta_m$. As we increase t, such band is spread over the region (-4t,4t), and very close to -4t the bands resemble parabolic dispersion. The phase transition occurs when -4t compensates \Delta_m and then the lowest energy levels are below the chemical potential.
Now, the non-trivial sectors of the Hilbert space must include the effects of $e$ particles or twisted boundary conditions as it follows from the representation of $X_{l}$ discussed in Eq.~\eqref{eq:DualityRelationX}. In general, the hopping part of the Hamiltonian is given by
\begin{IEEEeqnarray}{rCl}
    H_{t} & = & -t\sum_{\langle p,p' \rangle} e^{iA_{pp'}}\left( b_{p}^{\dagger}b_{p'} + b_{p'}^{\dagger}b_{p} \right) \\
    & = & -\dfrac{t}{2}\sum_{\langle p,p' \rangle} e^{iA_{pp'}} \left( X_{p}X_{p'}+Y_{p}Y_{p'} \right) \ ,
    \label{eq:HamBHbranchcuts}
\end{IEEEeqnarray}
where the effect of $A_{pp'}$ is just a change of sign in the hopping amplitude for the plaquettes adjacent to a branch cut.

% The remarkable difference is that the dual Ising magnet is able to adapt the phase and recover the random phase of the boson wave function far away from the branch cut. The phase itself doesn't have any physical meaning, but its behavior in both models is an alternative indicator of the physics in each phase/model.

% 2 - Including topology and $e$ particles

\subsection{Half-vortices and $e$ particle confinement}
\label{sec:Sec3A}

Let us now describe the energy cost of $e$ particle pairs as a function of their separation. As discussed in the previous section, such pair is connected by a branch cut. Contrary to the strong linear confinement of the pure $\mathbb{Z}_{2}$ model, $e$ particles produce a non-local modification of the $m$ particle superfluid by binding a half-vortex, as it is schematically shown in Fig.~\ref{fig:Fig4}b), and the required energy to separate a pair of $e$ particles follows a 2D Coulomb law, scaling only logarithmically with distance.

To demonstrate these properties, we use a Ginzburg-Landau description of the superfluid, which is valid at long-wavelengths whenever the system has a weak local deviation from the ground state. The Ginzburg-Landau energy functional can be expressed in terms of the complex superfluid order parameter $\braket{b}\sim\braket{X}+i\braket{Y}=\rho_{0}e^{i\phi}$. The long-wavelength limit allows to neglect higher-order spatial variations of the superfluid amplitude, leading to the following simplified energy functional: 
\begin{equation}
    E[\phi] = \int d^{2}x \dfrac{\rho_{s}}{2}\left( \vec{\nabla}\phi - \vec{A} \right)^{2} \ .
    \label{eq:PurePhaseStiffnessModel}
\end{equation}
Here $\vec{A}$ is the static non-dynamical vector potential that accounts for the branch cuts associated to $e$ particles and twisted boundary conditions, %$\rho_{s}=\rho_{0}^{2}/m$ 
$\rho_{s}$ is the superfluid phase stiffness %$m=(2ta^{2})^{-1}$, $a$ is the lattice spacing and lengths are given in units of $a$.
and lengths are given in units of the lattice constant. A more detailed description including the variations of the amplitude of the superfluid order parameter would regularize short distance divergences associated with the finite size of the vortex core, but we will be accounting for this here by simply introducing a cutoff by hand. 

Now consider some array of an even number of $e$ particles at vertices designated by locations $\vec{r}_{i}$. To compute the order parameter that minimizes the energy in the presence of such particles, we employ a familiar trick from the mapping of the XY model onto a Coulomb gas~\cite{LevitovNotes,BarNotes,kardar2007statistical}. Namely, we choose a smooth gauge in which the vector potential $\vec{A}$ absorbs all the purely transverse part of the superfluid current, and the phase $\phi$ is a smooth strictly continuous function containing the longitudinal part. Specifically, we choose the Coulomb gauge, $\vec{\nabla}\cdot\vec{A}=0$. To correctly reproduce the lattice Gauss law, the vector potential must satisfy
\begin{equation}
    \vec{\nabla}\times\vec{A} = \hat{z}\sum_{\mathrm{all} \ i} 2\pi v_{i}\delta(\vec{r}-\vec{r}_{i}) \ , \quad v_{i} = \pm\dfrac{1}{2}, \pm\dfrac{3}{2},\dots
    \label{eq:CurlA}
\end{equation}
The parameter $v_{i}$ is quantized to a half-integer, unlike the usual vortices where it is quantized to an integer. Therefore $e$ particles become sources of $\pi$ or half-vortices, which is essentially the mechanism for vortex fractionalization in an anyon superfluid first pointed by Kivelson in Ref.~[\onlinecite{kivelson1989statistics}]. The Coulomb gauge allows to separate the energy cost into singular and smooth parts without a crossed term, as follows
\begin{equation}
    E[\phi] = \int d^{2}x \dfrac{\rho_{s}}{2}\left( \left( \vec{\nabla}\phi \right)^{2} + \vec{A}^{2} \right) \ .
\end{equation}
The energy minimization is accomplished by setting $\phi$ constant while $\vec{A}$ is solved from Eq.~\eqref{eq:CurlA} giving
\begin{equation}
    \vec{A} = \vec{\nabla}\times\left( \chi \hat{z} \right) \ , \quad \chi = -\sum_{\mathrm{all} \ i}v_{i}\ln\left| \vec{r}-\vec{r}_{i} \right| \ .
\end{equation}
This vector potential leads to the energy functional of a 2D Coulomb gas
%\noteOP{Factor 2 added}
\begin{equation}
    E[\phi] = \sum_{\mathrm{all} \ i} \left( E_{c}(v_{i}) - \pi\rho_{s}\sum_{j>i} v_{i}v_{j}\ln\left| \vec{r}_{i}-\vec{r}_{j} \right| \right) \ ,
\end{equation}
where we have assumed that there is no net vorticity ($\sum_{i}v_{i}=0$) and we have implicitly added the self-interaction of vortices to the vortex core energy $E_{c}(v_{i})$. This manipulation is needed in the current model because there is a UV divergent vortex self-energy stemming from the fact that we are assuming point-like vortex cores, but such self-interaction would be finite in more realistic models in which the vortex cores are not point-like. In the special case of only two $e$ particles, they bind half-vortices of opposite vorticity and the energy cost to split them grows logarithmically with their distance
\begin{equation}
    E = 2E_{c} + \dfrac{\pi\rho_{s}}{2}\ln\left| \vec{r}_{1}-\vec{r}_{2} \right| \ .
\end{equation}

We conclude that $e$ particles are marginally deconfined in the $m$ particle superfluid phase, attracting each other with a logarithmic potential characteristic of the 2D Coulomb law. This is in sharp contrast to both their strong linear confinement in the confined phase of $\mathbb{Z}_{2}$ lattice gauge theory and the full deconfinement in the TC phase. Moreover, in this case the Coulomb phase remains separated by a BKT phase transition~\cite{kardar2007statistical} from the TC vacuum even at finite temperature. This contrast with the confined phase of $\mathbb{Z}_{2}$ lattice gauge theory, which is smoothly connected to the TC at finite temperature~\cite{fradkin2013field}. In this sense, one could say that the superfluid of $m$ particles is a more robust kind of marginal topological order with regard to temperature fluctuations.

Now it is conceivable to have a different phase in which $m$ particles do not Bose condense as individual particles but rather form a Bose condensate of molecule-like pairs. Such state would be characterized by a boson pair-order parameter with finite expectation $\braket{b^{2}}\sim \rho_{2}e^{i\phi_{2}}$, while the expectation value of the single boson remains zero, $\braket{b}=0$. For example, this could be engineered by taking the Hamiltonian from Eqs.~\eqref{eq:XXZmodel1}-\eqref{eq:XXZmodel2} and considering a hopping term in which there is still a gap for single boson excitations, namely $t<t_{c}=\Delta_{m}/4$, while adding a sufficiently strong boson attraction so that the two-boson bound state has negative energy making favorable for boson molecules to spontaneously proliferate in the vacuum. One would still need to ensure that the resulting bosonic molecules form a simple Bose condensate, which should occur if molecules have an effectively repulsive interaction so that upon boson pair-proliferation the system does not jump into the fully packed boson state, phase separates or orders in a crystalline or any other fashion, but as a matter of principle there should be no problem with engineering this situation in a microscopic Hamiltonian. In such $m$-pair condensate the Ginzburg-Landau energy functional would be
\begin{equation}
    E[\phi] = \int d^{2}x \dfrac{\rho_{s}}{2}\left( \vec{\nabla}\phi_{2} - 2\vec{A} \right)^{2} \ .
\end{equation}
The order parameter carries twice the charge and therefore sees the effective flux of $e$ particles as a $2\pi$ flux. These fluxes have no physical consequence on long-distance behavior of the superfluid order parameter and in particular are not required to bind a superfluid vortex. In this case the $e$ particle survives as a fully deconfined particle with a finite energy cost.

\subsection{$m$ particle superfluid in a torus}

We will now discuss the interplay of the $m$ particle Bose-condensation and the energy splitting of topological sectors with non-trivial loop operators in the torus. The sectors of the Hilbert space with non-trivial loop operators, $T_{x,y}=-1$, are mapped into antiperiodic boundary conditions as discussed in Sec.~\ref{sec:Sec2B}. In Sec.~\ref{sec:Sec2C} we saw that in the pure $\mathbb{Z}_{2}$ gauge theory such branch cut would induce an order $L_{x,y}$ energy cost in the state with twisted boundary conditions. This can be physically understood from the fact that the dual Ising magnet only has two low-energy orientations and therefore lacks any smooth way to heal away from such cut. In contrast, the superfluid or XY ferromagnet can smoothly adapt the boson phase so that the spins adjacent to the non-contractible branch cut are antiparallel and the rest twist smoothly away from the branch cut, leading just to a constant energy cost in the thermodynamic limit as we will estimate below. This comparison is depicted in Fig.~\ref{fig:Fig3}. 

% Ginzburg-Landau approach to prove previous features about branch cuts.

We can estimate the energy cost from such deformation of the superfluid ground state from the Ginzburg-Landau description of Sec.~\ref{sec:Sec3A}. It is convenient to employ a smooth gauge where the vector potential is spread over the torus, instead of the one described in Eq.~\eqref{eq:HamBHbranchcuts} that is localized in the single line of defective hopping bonds. These two gauges will be equivalent as long as their Wilson loop integrals around the non-contractible loops $\gamma_{\mathrm{nc}}$, as the one depicted in Fig.~\ref{fig:Fig3}, are identical. The lattice version of this condition is
\begin{equation}
    \prod_{(p,p')\in \gamma_{\mathrm{nc}}} e^{iA_{pp'}} = \prod_{(p,p')\in \gamma_{\mathrm{nc}}} e^{iA_{pp'}'} \ .
\end{equation}
For instance, considering $(T_{x},T_{y})=(-1,1)$ we use the vector potential $\vec{A}=(\pi/L_{x})\hat{x}$, and similarly for other cases. As a result, we find that the energy difference between the lowest energy states with twisted boundary conditions, measured relative to the absolute ground state with periodic boundary conditions, is given by:
\begin{IEEEeqnarray}{rCl}
    \Delta E\big|_{(T_{x},T_{y})=(-1,1)} & = & \dfrac{\rho_{s}\pi^{2}}{2}\dfrac{L_{y}}{L_{x}} \ , \label{eq:EnTwist1} \\
    \Delta E\big|_{(T_{x},T_{y})=(1,-1)} & = & \dfrac{\rho_{s}\pi^{2}}{2}\dfrac{L_{x}}{L_{y}} \ , \\
    \Delta E\big|_{(T_{x},T_{y})=(-1,-1)} & = & \dfrac{\rho_{s}\pi^{2}}{2}\dfrac{L_{x}^{2}+L_{y}^{2}}{L_{x}L_{y}} \ . \label{eq:EnTwist3}
\end{IEEEeqnarray}
Therefore, these sectors are split from the ground state by a finite energy cost that depends only on the aspect ratio of the torus. The ground states in these sectors are analogues of states where the supercurrent around the torus is associated with trapping a half-vortex in the inner loop of the torus. The results in Eqs.~\eqref{eq:EnTwist1}-\eqref{eq:EnTwist3} are in sharp contrast with both the energy splitting of the TC ground states, which decreases exponentially with system size~\cite{fradkin2013field,wen2004quantum}, and the confined phase of $\mathbb{Z}_{2}$ lattice gauge theory, where the splitting grows linearly with system size. 

On the other hand, we can consider again the scenario of a pair-condensate of $m$ particles, without single $m$ boson condensation. In this case the sectors with twisted boundary conditions $T_{x,y}=-1$ do not need to induce a super-flow around the torus in their respective low energy configurations, which remain degenerate in the thermodynamic limit with the vacuum at $T_{x,y}=1$. Therefore, apart from their distinction regarding the U(1) symmetry, the superfluid $m$-pair condensed state has the same topological order as the conventional TC.

\section{Toric Code and Fermionic $\mathbb{Z}_{2}$ Lattice Gauge Theory}
\label{sec:TC-FermionicZ2}

In Sec.~\ref{sec:BoseZ2} we reviewed how Hamiltonians containing pair fluctuations of $m$ particles and frozen $e$ particles lead to the classic structure of a $\mathbb{Z}_{2}$ lattice gauge theory. The states that one can access via this construction are limited by the bosonic nature of $m$ particles. An alternative kind of $\mathbb{Z}_{2}$ lattice gauge theory arises from the TC if one instead only allows for pair fluctuations of $\varepsilon$ particles. Because $\varepsilon$ particles have fermionic statistics among themselves, this procedure can be regarded as a precise form of 2D bosonization of fermions, as recently proposed in Ref.~[\onlinecite{chen2018exact}]. Here we review this construction and extend it to include fluxes and twists of boundary conditions on a finite 2D torus. For a discussion on the extension of this construction to lattices with open boundaries see Ref.~[\onlinecite{RaoInti2020}].

\begin{figure} [!t]
    \includegraphics{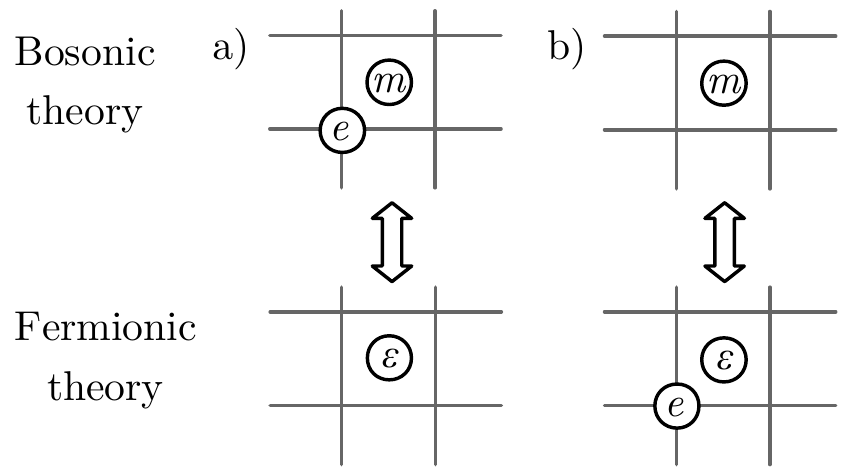}
    \caption{Depiction of the particles that are viewed as elementary building blocks in the bosonic and fermionic $\mathbb{Z}_{2}$ lattice gauge theories, showing the convention to label identical physical configurations in either theory. a) In the bosonic $\mathbb{Z}_{2}$ gauge theory, the elementary building blocks are the $e$ and $m$ particles, and the $\varepsilon$ fermion is viewed as a bound state of these. b) In the fermionic $\mathbb{Z}_{2}$ gauge theory, the elementary building blocks are the $\varepsilon$ fermion and the $e$ boson, whereas the $m$ particle is viewed as a composite.}
    \label{fig:FZ2-Fig1}
\end{figure}

In the bosonic description of Secs.~\ref{sec:ToricCodeZ2} and \ref{sec:BoseZ2}, we made a convention to view the $e$ and $m$ particle as the elementary particles providing the degrees of freedom that serve as building blocks to label all the states of the Hilbert space. In such choice the $\varepsilon$ particle appeared not as an elementary degree of freedom, but rather as a bound state of these two particles. The key idea behind the fermionic description is that we will take instead the $\varepsilon$ particle as one of the elementary particles together with the $e$ particle, and they will serve as the new building blocks to label all the states in the Hilbert space. In such convention, the $m$ particle will no longer be viewed as elementary but rather as a bound state of the $\varepsilon$ and $e$ particles. To do so, we begin by choosing a convention to uniquely specify the different particle configurations. 
For $\varepsilon$ particles we use a ``north-east'' charge-flux binding convention, in which each plaquette is paired with its north-east vertex as depicted in Fig.~\ref{fig:FZ2-Fig1}a). 
The representation of the TC vacuum excitations changes with respect to the bosonic $\mathbb{Z}_{2}$ lattice gauge theory. In the fermionic representation, we will say that $\varepsilon$ particles reside in the plaquettes, as shown in Fig.~\ref{fig:FZ2-Fig1}a), while $e$ particles still reside in the vertices. Since $\varepsilon$ and $e$ particles are ``hardcore'', their occupation of the plaquette and vertex sites are either 0 or 1. Isolated $m$ particles in this representation are now viewed as a bound state with an $\varepsilon$ particle to the north-east of an $e$ particle as shown in Fig.~\ref{fig:FZ2-Fig1}b). Isolated $e$ particles have the same form in both representations.

\begin{figure}[!t]
    \centering
    \includegraphics{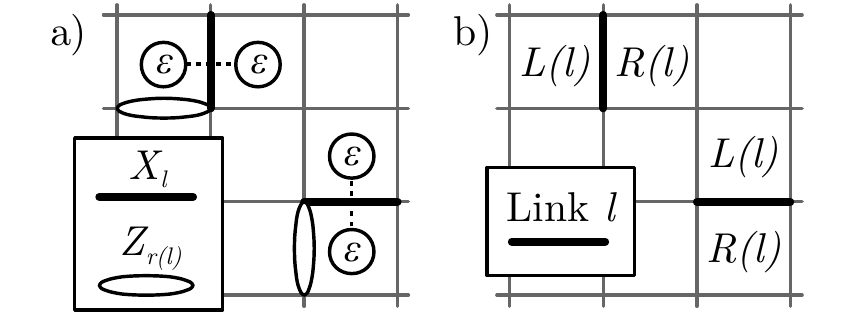}
    \caption{a) Pair fluctuation operators of $\varepsilon$ particles, defined in Eq.~\eqref{eq:EpsilonHop}, for the horizontal (left) and vertical (right) directions. Link labels $l$ and $r(l)$ specify in which links we apply $X$ and $Z$ operators. b) Notation for the neighbor plaquettes sharing a link $l$, which we introduce to perform the dual mapping of Eq.~\eqref{eq:FZ2-DualMapping2}.}
    \label{fig:FZ2-Fig2}
\end{figure}

The charge-flux binding convention implies a redefinition of operators measuring the parity of $e$ and $\varepsilon$ particle numbers relative to that of the bosonic theory in Section~\ref{sec:BoseZ2}. Now the $\varepsilon$ particle parity is measured by $G_{p}^{m}$, and the parity of $e$ particles is measured by $G_{p(v)}^{m}G_{v}^{e}$, where $p(v)$ is the plaquette to the north-east of vertex $v$ following the north-east flux convention. These local operators lead to the following $\varepsilon$ particle occupation of plaquettes
\begin{equation}
    n_{p} = \dfrac{1-G_{p}^{m}}{2} \ ,
    \label{eq:FL-Np}
\end{equation}
and $e$ particle occupation of vertices
\begin{equation}
    n_{v} = \dfrac{1-G_{p(v)}^{m}G_{v}^{e}}{2} \ .
    \label{eq:FL-Nv}
\end{equation}
The basic Hamiltonian that describes these elementary particles as gapped and frozen particles is given by
\begin{equation}
    H = \Delta_{\varepsilon} \sum_{p} n_{p} + \Delta_{e} \sum_{v} n_{v} \ ,
\end{equation}
where $\Delta_{\varepsilon},\Delta_{e}>0$. The $e$ particles are created in pairs by a string of $Z$ operators as in the bosonic theory, whereas $\varepsilon$ particles can be created by strings of pair fluctuation operators of the form
\begin{equation}
    S_{l} \equiv X_{l}Z_{r(l)} \ ,
    \label{eq:EpsilonHop}
\end{equation}
where $l$ is the link in between the pair fluctuation and $r(l)$ is defined in Fig.~\ref{fig:FZ2-Fig2}a). The local parity operators inherit the global constraints from Eq.~\eqref{eq:ParityConstraint} from the bosonic theory, which now becomes
\begin{IEEEeqnarray}{rCl}
    \prod_{v} G_{p(v)}^{m}G_{v}^{e} = 1 \ , & \qquad & \prod_{p} G_{p}^{m} = 1 \ ,
\end{IEEEeqnarray}
and imply again that the number of $e$ and $\varepsilon$ particles must be even in any state constructed in the torus.

\begin{figure}[!t]
    \centering
    \includegraphics{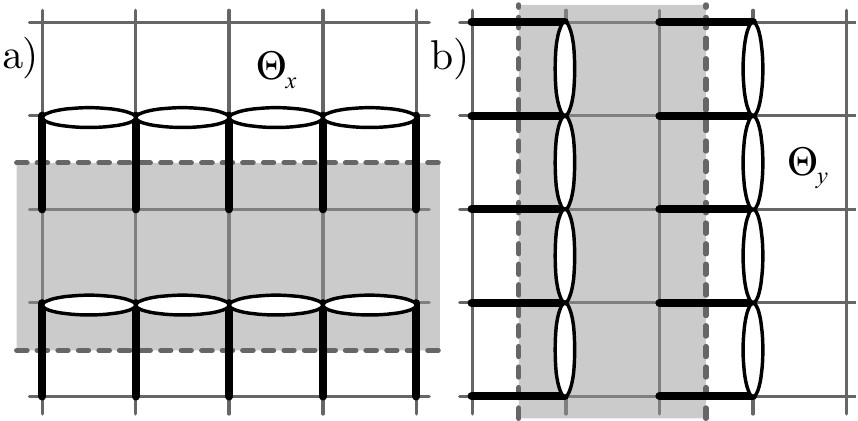}
    \caption{Non-contractible loop operators $\Theta_{x,y}$, defined in Eq.~\eqref{eq:FZ2-LoopOps}, as the boundaries of a multiply connected region $R$ (shaded in gray), which wraps the torus, where we apply the product of operators $G_{p(v)}^{m}G_{v}^{e}$ for every vertex $v$ contained in $R$.}
    \label{fig:FZ2-Fig4}
\end{figure}

To build the $\mathbb{Z}_{2}$ gauge structure, we only allow terms in the Hamiltonian that produce $\varepsilon$ particle pair fluctuations while keeping $e$ particles strictly frozen. Namely the Hamiltonian must be constructed out of local operators that commute with all operators $n_v$ from Eq.~\eqref{eq:FL-Nv} for every vertex $v$. This local conservation law plays the role of a new type of Gauss law~\cite{chen2018exact}. The complete basis for the algebra of local operators that satisfies these conditions consists of the operators $\left\{ n_{p}, S_{l} \right\}$. There are in addition two non-local t'Hooft loop operators that can be viewed as the operators associated with the boundary of the new Gauss law defined by $n_v$. To identify them, we consider a region $R$ that wraps the torus over one of the non-contractible directions. This region has two disconnected boundaries that form non-contractible loops around the torus, as depicted in Fig.~\ref{fig:FZ2-Fig4}. Namely, the product of $G_{p(v)}^{m}G_{v}^{e}$ over all the vertices $v$ contained inside $R$ can be written as the product of two of the following loop operators:
\begin{equation}
    \Theta_{x,y} = -\left( \prod_{ \ l \ \in \ \gamma_{x,y}} S_{l} \right) \left( \prod_{\ p \ \in \ \gamma_{x,y}} G_{p}^{m} \right) \ ,
    \label{eq:FZ2-LoopOps}
\end{equation}
where $\gamma_{x,y}$ is a non-contractible loop around the torus and the links $l\in\gamma_{x,y}$ and plaquettes $p\in\gamma_{x,y}$ that enter into the products in Eq.~\eqref{eq:FZ2-LoopOps} are illustrated in Fig.~\ref{fig:FZ2-Fig4}. The additional minus sign in Eq.~\eqref{eq:FZ2-LoopOps} is introduced for future notational convenience. Note that these operators commute with each other:
\begin{equation}
    \left[ \Theta_{x}, \Theta_{y} \right] = 0 \ .
\end{equation}

The topology of the region $R$ is essential to identify the global loop operators $\Theta_{x,y}$. In a simply connected region, the analogous loop operator at its boundary would be identical to the product of the parities of $e$ particles in its interior, and therefore it would not be an algebraically independent operator. On the contrary, the multiply connected region $R$ has two disconnected boundaries, and this leads to the non-contractible loop operators from Eq.~\eqref{eq:FZ2-LoopOps} to be algebraically independent from the local parity operator of $e$ particles, and therefore must be separately specified. These operators are the analogues of the $T_{x,y}$ operators defined in the bosonic case in Eq.~\eqref{eq:NCLoopOp}. Importantly, in spite of being algebraically independent from Eq.~\eqref{eq:FL-Nv}, the demand that the Hamiltonian commutes with every $n_{v}$ and that it is itself a sum of local operators, implies that the Hamiltonian must also commute with $\Theta_{x,y}$. Therefore, we conclude that in a 2D torus the Hilbert space with the fermionic $\mathbb{Z}_{2}$ gauge structure, splits into sectors labeled by the operators $\left\{ n_{v},\Theta_{x},\Theta_{y} \right\}$. This is the fermionic counterpart to the gauge structure described for bosons in Sec.~\ref{sec:Sec2B}.

\begin{figure}[!t]
    \centering
    \includegraphics{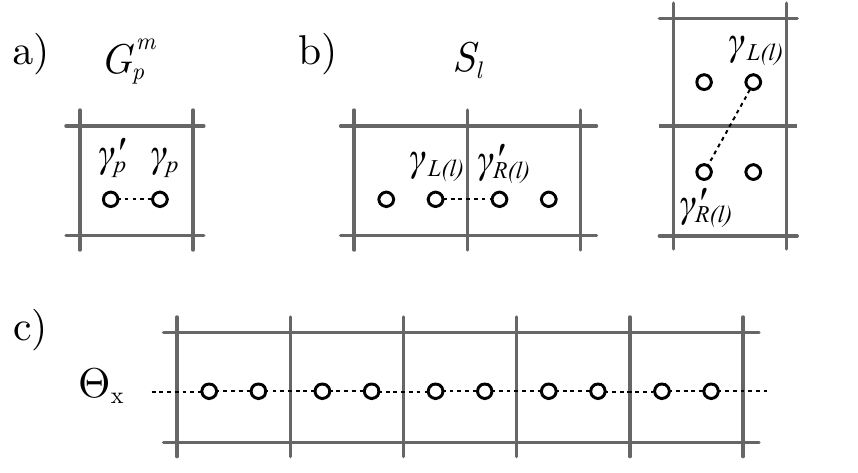}
    \caption{Dual mapping of the fermionic $\mathbb{Z}_{2}$ lattice gauge theory to Majorana fermion operators, defined in Eqs.~\eqref{eq:FZ2-DualMapping1}-\eqref{eq:FZ2-DualMapping2}. a) The plaquette operator $G_{p}^{m}$ creates a Majorana pair $\left\{ \gamma_{p},\gamma_{p}' \right\}$ at the plaquette $p$. b) The hopping operator $S_{l}$ creates Majorana fermions at the plaquettes sharing the link $l$, specifically a $\gamma_{p}$ at the plaquette $L(l)$ and a $\gamma_{p}'$ at the plaquette $R(l)$. c) The annihilation of a pair of Majorana fermions along a non-contractible loop in the $x,y$ direction of the torus is equivalent to the loop operator $\Theta_{x,y}$ defined in Eq.~\eqref{eq:FZ2-LoopOps}.}
    \label{fig:FZ2-FigMajoranas}
\end{figure}

Now let us discuss the dynamics of $\varepsilon$ particles within each of these subspaces labeled by $\left\{ n_{v}, \Theta_{x}, \Theta_{y} \right\}$. Remarkably, these are \emph{local} fermion spaces. Namely, there is an exact one-to-one correspondence between operators and states from the Hilbert subspace defined by $\left\{ n_{v}, \Theta_{x}, \Theta_{y} \right\}$ and the parity-even Hilbert space of spinless fermions residing on the plaquettes of the lattice. This mapping is strictly local in the sense that any local operator that commutes with $n_{v}$, and hence does not mix Hilbert subspaces with different $\left\{ n_{v},\Theta_{x},\Theta_{y} \right\}$, will map onto a local parity-even fermionic operator. We will now state the explicit mapping for the trivial sector without $e$ particles ($n_{v}=0$) and $\Theta_{x,y}=1$, extending the results of Ref.~[\onlinecite{chen2018exact}] to a finite lattice with periodic boundary conditions.

\begin{figure}[!t]
    \centering
    \includegraphics{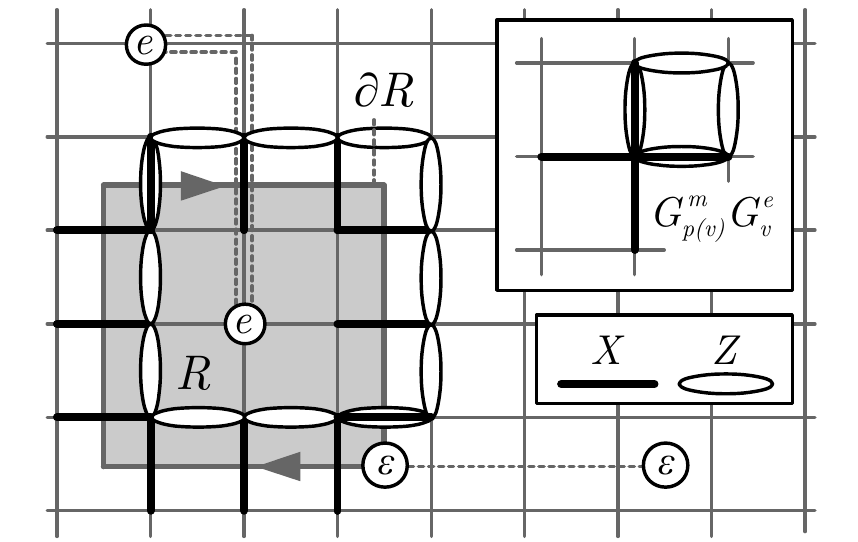}
    \caption{Braiding law for $\varepsilon$ particles in the fermionic $\mathbb{Z}_{2}$ lattice gauge theory. Similar to the bosonic case, the braiding of an $\varepsilon$ particle along the boundary $\partial R$ results in a global sign in the final state. This sign is equivalent to the $e$ particle parity inside $R$, which is the product of $G_{p(v)}^{m}G_{v}^{e}$ for all $v\in R$.}
    \label{fig:FZ2-Fig3}
\end{figure}

Because the local dimension of the Hilbert space associated with each plaquette is $2$ ($G_{p}^{m}=\pm1$), we introduce a dual complex fermion mode operator $\left\{ c_{p},c_{p}^{\dagger} \right\}$ or, equivalently, two dual Majorana fermion mode operators per plaquette $\left\{ \gamma_{p},\gamma_{p}' \right\}$:
\begin{equation}
    c_{p} = \dfrac{\gamma_{p}+i\gamma_{p}'}{2} \ . \label{eq:Majoranas}
\end{equation}
The local operators $\left\{ G_{p}^{m},S_{l} \right\}$ are mapped into dual Majorana operators as follows (see Fig.~\ref{fig:FZ2-FigMajoranas})
\begin{IEEEeqnarray}{rCl}
    G_{p}^{m} \ & \Longleftrightarrow & \ i\gamma_{p}'\gamma_{p} \label{eq:FZ2-DualMapping1} \\
    S_{l} \ & \Longleftrightarrow & \ i\gamma_{L(l)}\gamma_{R(l)}' \label{eq:FZ2-DualMapping2}
\end{IEEEeqnarray}
Here, $\{ L(l) , R(l) \}$ denote neighbor plaquettes sharing the link $l$, following the convention depicted in Fig.~\ref{fig:FZ2-Fig2}b). In order to extend this dual representation to the rest of Hilbert subspaces, we note that the parity operator $\Pi_{e}$, which measures the parity of the total number of $e$ particles inside a simply connected region of the torus $R$, reads as:
\begin{equation}
    \Pi_{e} \equiv \prod_{v\in R} G_{p(v)}^{m}G_{v}^{e} = \prod_{v\in R}(-1)^{n_{v}} \ ,
    \label{eq:Parity-e-FZ2}
\end{equation}
The above operator can be viewed as closed loop operator acting only on the spins residing in the boundary of the region $R$ and not in its interior, as depicted in Fig.~\ref{fig:FZ2-Fig3}. Such boundary loop operator can be interpreted as the operator that transports the $\varepsilon$ fermions on closed contractible loops, and thus measures the parity of the $e$ particles inside the loop due to the their non-local statistical interaction. However, the representation from Eqs.~\eqref{eq:FZ2-DualMapping1}-\eqref{eq:FZ2-DualMapping2} automatically implies that $\Pi_{e}=1$ and $\Theta_{x,y}=1$, which restricts the validity of such dual representation to the subspace without $e$ particles, $n_{v}=0$ for all $v$, and with trivial loop eigenvalues $\Theta_{x,y}=1$. To find the dual representation of the remaining subspaces, we follow a similar construction to Sec.~\ref{sec:ToricCodeZ2}, namely we associate a string of links connecting pairs of $e$ particles, so that each $e$ particle has a unique string emanating from it. Each of these strings is a ``branch cut'' where we will twist the representation of pair-creation operators from Eq.~\eqref{eq:FZ2-DualMapping2}, by adding a vector potential $A_{l}$, so that there is an extra minus sign in Eq.~\eqref{eq:FZ2-DualMapping2} when the link $l$ belongs to this branch cut. Mathematically,
\begin{equation}
    S_{l} \ \Longleftrightarrow \ e^{iA_{l}}\left( i\gamma_{L(l)}\gamma_{R(l)}' \right)
\end{equation}
where $A_{l}=\pi$ if the link $l$ belongs to the branch cut and $0$ otherwise. Similarly, the operators $\Theta_{x,y}$ from Eq.~\eqref{eq:FZ2-LoopOps} can be viewed as $\varepsilon$ fermion transport over closed non-contractible loops around the torus as it is depicted in Fig.~\ref{fig:FZ2-FigMajoranas}. To correctly represent subspaces with $\Theta_{x,y}=-1$, we introduce non-contractible branch cuts across the torus. In other words, the Hilbert spaces with $\Theta_{x,y}=1$ and $-1$ correspond respectively to periodic and anti-periodic boundary conditions for the $\varepsilon$ fermions around the torus. Following this recipe, we have an \emph{exact} representation of any subspace $\left\{ n_{v},\Theta_{x},\Theta_{y} \right\}$ as a space of fermions coupled to non-dynamic $\pi$-flux tubes and twisted boundary conditions on the torus.

This dual mapping allows to explicitly construct a large class of exactly solvable models, one for each free fermion bilinear Hamiltonian. The classic honeycomb model introduced by Kitaev~\cite{kitaev2006anyons} can in fact be viewed as an example of these Hamiltonians, as shown in Ref.~[\onlinecite{chen2018exact}]. A particularly appealing feature of the current construction is that it bypasses the need to enlarge the local physical Hilbert space, as it is customarily done in parton descriptions. As before, the only unphysical symmetry of the dual Hamiltonian is the global fermion parity, which should be viewed as a global constraint 
%\color{blue}
that can be easily enforced by only taking the states with an even number of $\varepsilon$ particles as physical. 
%\color{black}
In the absence of further symmetries, the gapped phases of free fermions are classified by the spectral Chern number $\nu\in\mathbb{Z}$, which is half of the ordinary Chern number in the presence of charge conservation and counts the number of chiral Majorana fermion modes at the edge~\cite{kitaev2006anyons,schnyder2008classification,kitaev2009periodic,schnyder2009classification,ryu2010topological}. The nature of $e$ particles in the presence of such fermionic states can be profoundly modified. The reason is that the $\pi$-flux is non-trivially dressed by the surrounding fermionic liquid. In particular for the case $\nu=1$, where $\varepsilon$ particles form a $p+ip$-type paired state, the $e$ particle would carry a Majorana zero mode in its core~\cite{read2000paired}. Such dressed composite is also referred to as the $\sigma$ particle. Interestingly, the bulk topological properties of such dressed $e$ particle only depend on $\nu$ modulo $16$, as demonstrated by Kitaev~\cite{kitaev2006anyons}. Upon adding symmetries, a far richer and finer refinement of the landscape of possible phases appears, and the present construction allows to write exactly solvable models for any of such phases obtained from free fermion Hamiltonians. For example, in Ref.~[\onlinecite{RaoInti2020}], the current formalism is employed to enlarge the $\nu$  $\mathrm{mod}(16)$ classification of Kitaev~\cite{kitaev2006anyons}, by accounting for translational symmetry. For a recent review on topological phases with symmetry see e.g. Ref.~[\onlinecite{chiu2016classification}]. We will demonstrate in the next section that a rich set of properties of the dressed $e$ particle appears when it is embedded in a gapless Fermi liquid of $\varepsilon$ particles.

\section{$\mathbb{Z}_{2}$ anyon Fermi liquids}
\label{sec:Z2FermiLiquids}
In this section we discuss how to endow the $\varepsilon$ fermions of the TC with a global U(1) symmetry and explore the resulting Fermi liquid state that follows from a chemical potential driven phase transition into a state with a finite fermion density. One of our goals is to determine whether the descendants of $e$ particles which act as thin solenoids of $\pi$-flux for these fermions, remain as fully deconfined finite energy particles in spite of coupling non-locally to a surrounding fluid of $\varepsilon$ fermions. Remarkably, we will see that these particles indeed remain fully deconfined in the presence of the Fermi sea in spite of producing long-ranged power law disturbances to the Fermi fluid. This property is relevant to phases with emergent fermions that carry microscopically conserved U(1) quantum numbers and emergent $\mathbb{Z}_{2}$ gauge charges and form a state with a Fermi surface, such as the orthogonal metal~\cite{nandkishore2012orthogonal}, and our calculations can be transferred directly to understand the deconfinement of the vison excitation~\cite{senthil2000z} in such phases. For other constructions of $\mathbb{Z}_{2}$ gauge theories with finite fermion density see e.g. Ref.~[\onlinecite{prosko2017simple}].

In analogy to the ordinary $\mathbb{Z}_{2}$ case, we construct Hamiltonians that commute with the total number of $\varepsilon$ particles, which following the discussion from Sec.~\ref{sec:TC-FermionicZ2} can be defined as:
\begin{equation}
    N_{\varepsilon} = \sum_{p}n_{p} = \sum_{p} \left( \dfrac{1-G_{p}^{m}}{2} \right) \ .
\end{equation}
The simplest non-trivial operator satisfying this requirement is the complex fermion hopping between nearest-neighbor plaquettes, $\left\{ p_{1},p_{2} \right\}$. In terms of Majorana operators defined in Eq.~\eqref{eq:Majoranas}, it reads as:
\begin{IEEEeqnarray}{rCl}
    c_{p_{1}}^{\dagger}c_{p_{2}} & = & \left( \dfrac{\gamma_{p_{1}}-i\gamma_{p_{1}}'}{2} \right) \left( \dfrac{\gamma_{p_{2}}+i\gamma_{p_{2}}'}{2} \right) \\
    & = & \dfrac{\gamma_{p_{1}}\gamma_{p_{2}}+\gamma_{p_{1}}'\gamma_{p_{2}}' + i\left( \gamma_{p_{1}}\gamma_{p_{2}}' - \gamma_{p_{1}}'\gamma_{p_{2}} \right)}{4} \ . \qquad
\end{IEEEeqnarray}
%\color{blue}
By employing the dual representation from Eqs.~\eqref{eq:FZ2-DualMapping1}-\eqref{eq:FZ2-DualMapping2}, we can relate fermion operators $\left\{ c_{p},c_{p}^{\dagger} \right\}$ with the $\varepsilon$ pair fluctuation operator Eq.~\eqref{eq:EpsilonHop} and the local $\varepsilon$ particle occupation from Eq.~\eqref{eq:FL-Np} as follows:
\begin{IEEEeqnarray}{rCl}
    c_{L(l)}^{\dagger}c_{R(l)} \ & \Longleftrightarrow & \ n_{L(l)} \ S_{l} \ n_{R(l)} \ , \label{eq:FZ2-DualHopping1} \\
    c_{R(l)}^{\dagger}c_{L(l)} \ & \Longleftrightarrow & \ n_{R(l)} \ S_{l} \ n_{L(l)} \ , \label{eq:FZ2-DualHopping2}
\end{IEEEeqnarray}
% Use the correspondence
% \begin{equation}
%    \gamma_{p} = c_{p} + c_{p}^{\dagger}
%    \gamma_{p}' = i ( c_{p}^{\dagger} - c_{p} )
% \end{equation}
where $L(l)$ and $R(l)$ are plaquettes adjacent to the link $l$ with the convention depicted in Fig.~\ref{fig:FZ2-Fig2}b), and $n_{p}$ and $S_{l}$ are the operators defined in Eq.~\eqref{eq:FL-Np} and \eqref{eq:EpsilonHop}, respectively. Notice the closely related structure to the bosonic theory described in Section~\ref{sec:BoseZ2}. Fermion hopping operators between further neighbor plaquettes can be constructed in analogous fashion. The representation of Eqs.~\eqref{eq:FZ2-DualHopping1}-\eqref{eq:FZ2-DualHopping2} corresponds to the trivial case without $e$ particles and periodic boundary conditions in the torus, $\Theta_{x,y}=1$. As in the bosonic case, the representation needs to include a vector potential associated with the $\pi$-fluxes carried by $e$ particles and antiperiodic boundary conditions, $\Theta_{x,y}=-1$, in the form of ``branch cuts'',
\begin{IEEEeqnarray}{rCl}
    e^{iA_{l}}c_{L(l)}^{\dagger}c_{R(l)} \ & \Longleftrightarrow & \ n_{L(l)} \ S_{l} \ n_{R(l)} \ , \label{eq:FZ2-GeneralDualHopping1} \\
    e^{iA_{l}}c_{R(l)}^{\dagger}c_{L(l)} \ & \Longleftrightarrow & \ n_{R(l)} \ S_{l} \ n_{L(l)} \ . \label{eq:FZ2-GeneralDualHopping2}
\end{IEEEeqnarray}
As a concrete example, we choose fermions residing on the plaquettes of a square lattice with only nearest-neighbor hoppings. The Hamiltonian expressed in terms of the microscopic spin degrees of freedom living on the links is,
\begin{IEEEeqnarray}{rCl}
    H & = & -\dfrac{t}{2}\sum_{\langle p_{1},p_{2} \rangle}S_{l}\left( 1 - G_{L(l)}^{m}G_{R(l)}^{m} \right) + \nonumber \\
    & & + \Delta_{\varepsilon}\sum_{\mathrm{all} \ p} \left( \dfrac{1-G_{p}^{m}}{2} \right) + \Delta_{e}\sum_{\mathrm{all} \ v}\left( \dfrac{1-G_{p(v)}^{m}G_{v}^{e}}{2} \right)  , \qquad
\end{IEEEeqnarray}
where $\Delta_{\varepsilon},\Delta_{e}>0$. In the subspace without $e$ particles and periodic boundary conditions, $\left\{ n_{v},\Theta_{x},\Theta_{y} \right\} = \left\{ 0,1,1 \right\}$, this Hamiltonian maps \emph{exactly} onto free fermions hopping in the square lattice,
\begin{equation}
    H = -t\sum_{\langle p_{1},p_{2} \rangle} \left(  c_{p_{1}}^{\dagger}c_{p_{2}} + c_{p_{2}}^{\dagger}c_{p_{1}} \right) + \Delta_{\varepsilon}\sum_{\mathrm{all} \ p}c_{p}^{\dagger}c_{p} \ .
    \label{eq:FreeFermionSquareLattice}
\end{equation}
Just as before, the only constraint that needs to be imposed in the fermionic representation for it to be in one-to-one correspondence with the states of the spin model is the global restriction to states with even fermion parity. This model will undergo a phase transition for $t=\Delta_{\varepsilon}/4$ from the vacuum of the ordinary TC with $N_{\varepsilon}=0$ to a state with a finite $\varepsilon$ fermion density and a Fermi surface that grows as $t/\Delta_{\varepsilon}$ increases in magnitude.
% Here we assume for the phase transition that the chemical potential lies at $\mu=0$. Then when t=0 the energy bands are Toric Code like, and they are flat and separated $\Delta_\varepsilon$. As we increase t, such band is spread over the region (-4t,4t), and very close to -4t the bands resemble parabolic dispersion. The phase transition occurs when -4t compensates \Delta_\varepsilon and then the lowest energy levels are below the chemical potential.

In the presence of $e$ particles and antiperiodic boundary conditions on the torus, $\Theta_{x,y}=-1$, the situation is completely analogous to that described in previous sections, namely, the Hamiltonian maps onto one in which $e$ particles act as static thin solenoids with $\pi$-flux, while the twists $\Theta_{x,y}=-1$ lead to antiperiodic boundary conditions in the torus. Therefore, omitting the constant energy associated with each $e$ particle ($\Delta_{e}$), the more general Hamiltonian is given by
\begin{IEEEeqnarray}{rCl}
    H & = & -t\sum_{\mathrm{all} \ l}\left( e^{iA_{l}}c_{L(l)}^{\dagger}c_{R(l)} + \mathrm{h.c.} \right) + \Delta_{\varepsilon}\sum_{\mathrm{all} \ p} c_{p}^{\dagger}c_{p} . \qquad
    \label{eq:FZ2-FullLatticeModel}
\end{IEEEeqnarray}

\subsection{Continuum limit in disk geometry}

To study the energy of the $e$ particle we will first consider the case in which the fermion density is sufficiently small so that a parabolic band dispersion can be employed to approximate the dispersion near the bottom of the band. This is not crucial for the validity of our conclusions, as we will demonstrate by direct calculations on the square lattice, but allows for a simplified treatment. We introduce a single $e$ particle in the form of an infinitesimally thin solenoid with $\pi$ flux located at the origin. Therefore, the Hamiltonian of the $\varepsilon$ particles is given by
\begin{equation}
    H = \dfrac{1}{2m_{\varepsilon}} (\vec{p}-\vec{A})^{2} \ , \qquad \vec{A} = \dfrac{\Phi}{2\pi r}\hat{\varphi} \ .
    \label{eq:FZ2-Disk-Analytic}
\end{equation}
Here $\Phi=\left\{ 0,\pi \right\}$ depending on whether the $e$ particle is present or not at the origin, $m_{\varepsilon}=1/(2ta^{2})$ and we have set the lattice constant $a=1$. We are omitting a constant energy term coming from the bottom of the band energy in the tight-binding model, given by $-4t$. As we will see, our calculation has a similar flavor to that of the scaling dimensions of monopole operators of Dirac fermions coupled to compact U(1) gauge fields~\cite{borokhov2003topological} which has been used to argue the stability of Dirac spin liquids in the limit of a large number of flavors~\cite{hermele2004stability}. However, the exact fermion duality we have described combined with the fact that $e$ particles are completely immobile in our ideal Hamiltonian, makes it clear that our calculation provides an essentially exact answer to the question of the deconfinement of $e$ particles in $\mathbb{Z}_{2}$ Fermi fluids of the $\varepsilon$ particles. We place the system in a disk of radius $R$, and impose hard wall boundary conditions, $\Psi(r=R)=0$, on the wavefunctions of the $\varepsilon$ particles. The single particle wavefunctions for this geometry can be expressed in terms of Bessel functions and are given by
\begin{IEEEeqnarray}{rCl}
    \Psi(\vec{r},\Phi=0) & = & \mathcal{C}_{0} J_{l}\left( |x_{l,n}| r/R \right) \ , \\
    \Psi(\vec{r},\Phi=\pi) & = & \mathcal{C}_{\pi} J_{|l-1/2|}\left( |x_{l-1/2,n}| r/R \right) \ ,
\end{IEEEeqnarray}
where $\mathcal{C}_{0},\mathcal{C}_{\pi}$ are normalization constants, $x_{l,n}$ is the n$th$ zero of the Bessel function $J_{l}(x)$, $n\in\mathbb{N}$ and $l\in\mathbb{Z}$. The energy levels in each flux configuration are given by
\begin{IEEEeqnarray}{rCl}
    E_{l,n}(\Phi=0) & = & %\dfrac{\hbar^{2}}{2m_{\varepsilon}R^{2}}
    \dfrac{x_{l,n}^{2}}{2m_{\varepsilon}R^{2}} \ , \label{eq:BesselEn0} \\
    E_{l,n}(\Phi=\pi) & = & %\dfrac{\hbar^{2}}{2m_{\varepsilon}R^{2}}
    \dfrac{x_{|l-1/2|,n}^{2}}{2m_{\varepsilon}R^{2}} \ . \label{eq:BesselEnPi}
\end{IEEEeqnarray}
All energy levels except $l=0$ for the $\Phi=0$ configuration are doubly degenerate.
Notice that we have excluded all possible solutions which are either non-normalizable or have divergent wavefunctions at the origin. One might think that the solutions that are normalizable but have divergent probability amplitude at the origin are physical. However, they can be discarded with a more detailed model in which the $e$ particle is taken as a hard wall solenoid with finite radius $R_{\mathrm{UV}}$, $\Psi(r=R_{\mathrm{UV}})=0$, and take the limit $R_{\mathrm{UV}}\rightarrow0$, where one recovers states with the same energies as those in Eqs.~\eqref{eq:BesselEn0}-\eqref{eq:BesselEnPi}. Notice that such limit does not eliminate the orbitals with finite amplitude at the origin, of the form $J_{0}(\sqrt{2m_{\varepsilon} E_{n,l}(0)}r)$, but mixes them with Bessel functions of second kind $Y_{0}(\sqrt{2m_{\varepsilon} E_{n,l}(0)} r)$, and the energies of the orbitals are identical to Eq.~\eqref{eq:BesselEn0} in the limit $R_{\mathrm{UV}}\rightarrow0$. We have also verified that our answers reproduce the direct calculation with the explicit lattice tight-binding model in the limit of small densities, as we will detail later on.

\begin{figure}[!t]
    \centering
    \includegraphics*{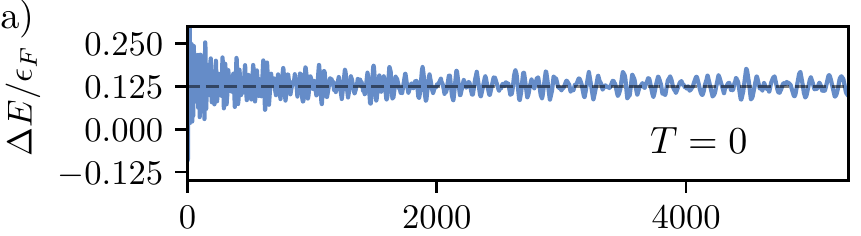}
    \includegraphics*{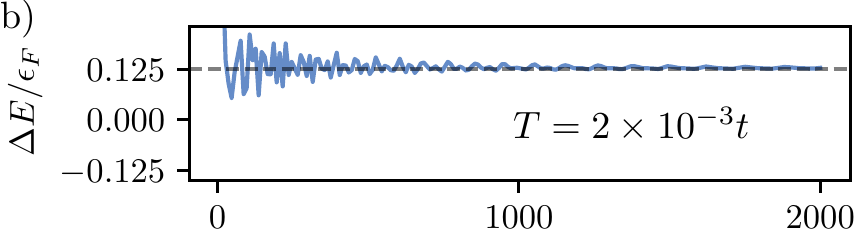}
    \includegraphics*{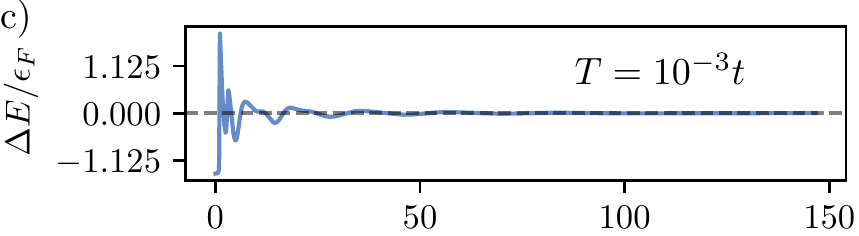}
    \includegraphics*{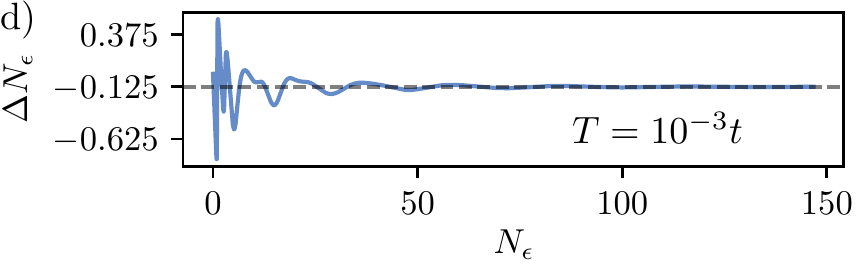}
    \caption{Energy and particle number differences between the state with and without $\pi$-flux in the continuum disk model from Eq.~\eqref{eq:FZ2-Disk-Analytic}. a),b) Energy difference fixing the $\varepsilon$ particle number at zero and finite temperature, respectively. c),d) Energy and particle number differences fixing the chemical potential $\mu$ at finite temperature, respectively.}
    \label{fig:FZ2-Fig5}
\end{figure}

In order to extract the thermodynamic limiting behavior it is convenient to introduce a chemical potential and a finite temperature. This leads to a Fermi-Dirac occupation, $n_{F}(E,\mu,T)$, for single particle states of energy $E$, and the following expressions for the total energy and particle number:
\begin{IEEEeqnarray}{rCl}
    n_{F}(E,\mu,T) & = & \dfrac{1}{\exp\left( (E-\mu)/(k_{B}T) \right) + 1} \ , \\
    N_{\varepsilon}(\Phi,\mu,T) & = & \sum_{i=1}^{\infty} n_{F}(E_{i},\mu,T) \ , \\
    E(\Phi,\mu,T) & = & \sum_{i=1}^{\infty} n_{F}(E_{i},\mu,T) E_{i}(\Phi) \ ,
\end{IEEEeqnarray}
where $i$ is a label for all the single particle states. To characterize the differences between ground states with and without flux in the thermodynamic limit, we must take the limits of $N_{\varepsilon}\rightarrow\infty$ and $T\rightarrow0$. The energy difference between the Fermi gas with and without flux ($\Phi=0,\pi$) and a common value of $N_{\varepsilon}$ at $T=0$ is shown in Fig.~\ref{fig:FZ2-Fig5}a). We find that the difference approaches a constant value, consistent with the complete energetic deconfinement of the $e$ particle ($\pi$-flux), given by
\begin{equation}
    \lim_{N_{\varepsilon}\rightarrow\infty} \left\{ E(N_{\varepsilon},\Phi=\pi) - E(N_{\varepsilon},\Phi=0) \right\}_{T=0} = \dfrac{\epsilon_{F}}{8} \ ,
    \label{eq:EnDiff-FixedN}
\end{equation}
where $\epsilon_{F}=2\pi n_{\varepsilon}/m_{\varepsilon}$ is the Fermi energy of the $\varepsilon$ particles, which equals the difference of the chemical potential and the energy of the band bottom for $T\rightarrow 0$, and $n_{\varepsilon}$ is the density of the $\varepsilon$ fermions. The results in Fig.~\ref{fig:FZ2-Fig5}a) show strong finite size fluctuations, although the limiting value is clear. One can mitigate such fluctuations by taking the zero temperature limit ($T\rightarrow0$) and large system size limit ($N_{\varepsilon}\rightarrow\infty$), while keeping the temperature larger than the finite size level spacing $\Delta\epsilon\sim\epsilon_{F}/\sqrt{N_{\varepsilon}}$, namely by keeping $\epsilon_{F}\gg T \gg \Delta\epsilon$. Fig.~\ref{fig:FZ2-Fig5}b) illustrates this idea by showing that at finite but small temperatures the system approaches the same limit as that in Eq.~\eqref{eq:EnDiff-FixedN} while avoiding the strong finite size fluctuations of the $T=0$ case. This figure has been obtained by adjusting the chemical potentials of the states with and without flux ($\Phi=0,\pi$) to ensure that we always compare systems the same total $N_{\varepsilon}$. Therefore, we conclude that there is a finite energy to add the $e$ particle onto the Fermi fluid of $\varepsilon$ particles given by Eq.~\eqref{eq:EnDiff-FixedN}.

\begin{figure}[!]
    \centering
    \includegraphics[scale=1.0]{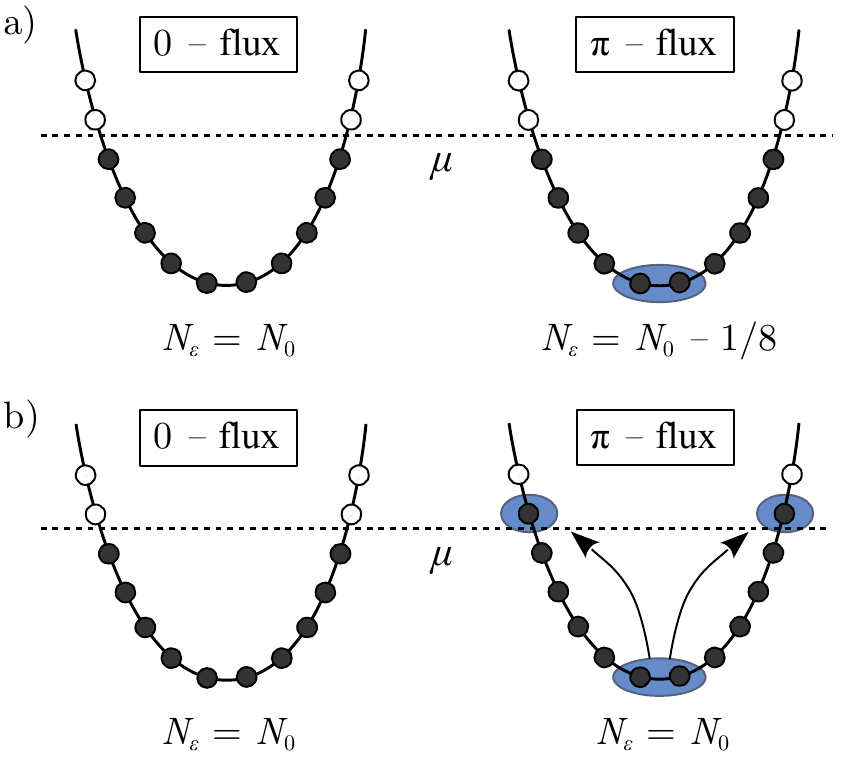}
    \caption{Depiction of the difference of states with and without an $e$ particle ($\pi$-flux). a) At fixed chemical potential, the $\pi$-flux is surrounded by a hole of $\varepsilon$ fermions, which arises by depleting states at the bottom of the band (shaded in blue). b) When the $\pi$-flux is inserted at fixed $\varepsilon$ particle number, the particles removed from the bottom of the band are added at the Fermi level.}
    \label{fig:Extra}
\end{figure}

To further understand the nature of the $e$ particle, we will now compare the difference of energies and particle numbers at fixed common chemical potentials between the states with and without flux. We again keep the temperature small but finite to avoid strong system size fluctuations and obtain the behavior in the limit $\epsilon_{F} \gg T \gg \Delta\epsilon$. Figs.~\ref{fig:FZ2-Fig5}c) and d) show that in this case we have the following limits
\begin{IEEEeqnarray}{rCl}
    \lim_{T\rightarrow0}\lim_{R\rightarrow\infty} \left\{ E(\mu,\Phi=\pi,T) - E(\mu,\Phi=0,T) \right\} & = & 0 \ , \qquad \\
    \lim_{T\rightarrow0}\lim_{R\rightarrow\infty} \left\{ N_{\varepsilon}(\mu,\Phi=\pi,T) - N_{\varepsilon}(\mu,\Phi=0,T) \right\} & = & -\dfrac{1}{8} \ . \qquad
\end{IEEEeqnarray}
The above equation demonstrates that the $\pi$-flux state carries a deficiency of $1/8$ of an $\varepsilon$ fermion relative to the state with no flux. The fact that the energy difference is $0$, can be understood by picturing that the fermion-hole that is created by inserting the flux is made from states near the bottom of the band which have nearly vanishing kinetic energy, and thus leading to vanishing energy difference as illustrated in Fig.~\ref{fig:Extra}. This also explains why the energy cost of inserting the flux at fixed $\varepsilon$ particle number in Eq.~\eqref{eq:EnDiff-FixedN} is exactly $\epsilon_{F}/8$, namely, because the $1/8$ fermions removed from the bottom of the band need to be accommodated at the Fermi level since lower energy states are Pauli-blocked, as illustrated in Fig.~\ref{fig:Extra}.

\begin{figure}[!t]
    \centering
    \includegraphics{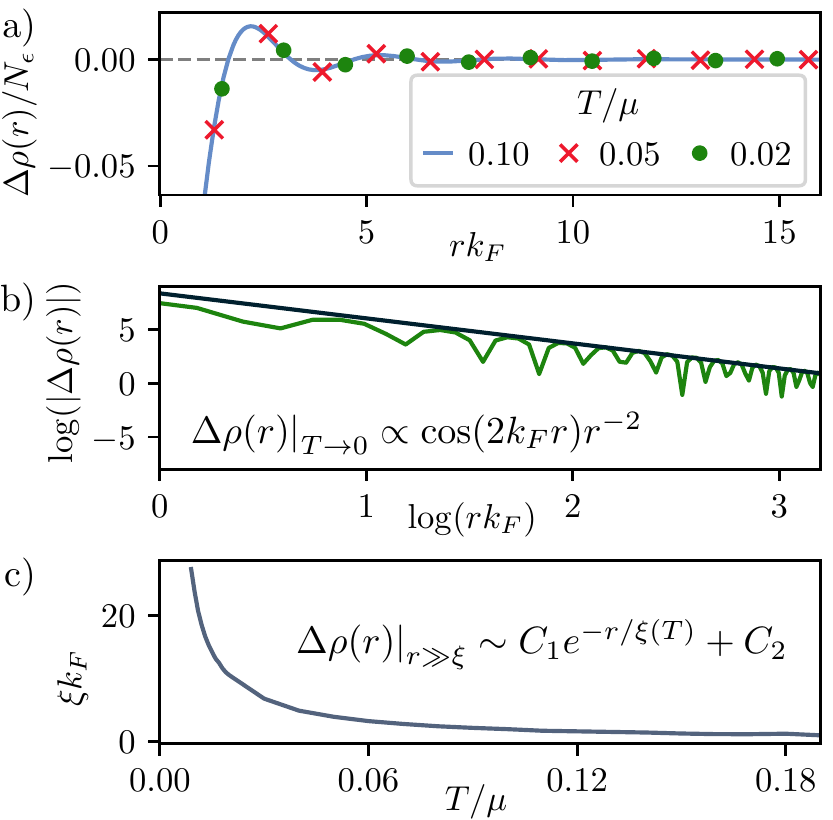}
    \caption{a) Local density difference between the $\pi$ and zero flux configurations of the $\varepsilon$ Fermi liquid at fixed chemical potential $\mu$. b) Near the $\pi$-flux core, the density perturbation displays a Friedel-like oscillatory decay with a power that approaches $\alpha=-2$ in the zero temperature limit. The plot shows the envelope fitting for $T/\mu=0.02$, for which $\alpha=-2.32$. The c) At finite temperatures, there exist a correlation length $\xi(T)$ above which the decay becomes exponential. Such correlation length diverges in the zero temperature limit. Results for the analytic approach from Eq.~\eqref{eq:FZ2-Disk-Analytic} with disk geometry.}
    \label{fig:Fig-11-FZ2}
\end{figure}

To study the spatial distribution of the screening cloud of $\varepsilon$ fermions surrounding the $e$ particle ($\pi$-flux), we compute the change of the local density of the $\varepsilon$ fluid, defined as
\begin{IEEEeqnarray}{rCl}
    \rho_{\varepsilon}(r,\Phi) & = & \sum_{i=1}^{\infty} n_{F}(E_{i}(\Phi),\mu,T) |\Psi_{i}(r,\Phi)|^{2} \ , \\
    \Delta\rho_{\varepsilon}(r) & = & \rho_{\varepsilon}(r,\pi)-\rho_{\varepsilon}(r,0) \ .  
\end{IEEEeqnarray}
Fig.~\ref{fig:Fig-11-FZ2}a) shows that the density is modified in the near vicinity of the flux. We find that the density fluctuations away from the core of the flux can be fit as follows:
\begin{IEEEeqnarray}{rClrl}
    \Delta\rho_\varepsilon(r) & \approx & A \cos\left( 2k_{F}r \right) r^{-\alpha(T)} & \quad \mathrm{for} \ \ & r\ll\xi(T) \ , \\
    \Delta\rho_{\varepsilon}(r) & \approx & B \cos\left( 2k_{F}r \right) e^{-r/\xi(T)} & \quad \mathrm{for} \ \ & \xi(T)\ll r \ll R \ , \qquad
\end{IEEEeqnarray}
as depicted in Fig.~\ref{fig:Fig-11-FZ2}b). Here $\xi(T)$ is a finite temperature correlation length that diverges as the temperature is lowered as shown in Fig.~\ref{fig:Fig-11-FZ2}c). This exponential decay at finite temperature is generic of Fermi liquids, because temperature acts as a relevant perturbation and there is no sharp phase transition separating them at finite temperature from the completely uncorrelated infinite temperature state with a finite (zero) correlation length. At zero temperature, we encounter that the $e$ particle ($\pi$-flux) induces an oscillatory power-law-decaying disturbance of the density of the liquid with a power that approaches $\alpha \approx 2$ as $T\rightarrow0$, and period $2k_F$ as it is the case of 2D Friedel oscillations~\cite{giuliani2005quantum}. This behavior is a hallmark of the presence of a sharp Fermi surface. The fact that the $e$ particle remains a finite energy excitation in spite of inducing such long-range disturbance on the surrounding liquid of $\varepsilon$ fermions can be understood by the vanishing cost of exciting particle-hole pairs infinitesimally close to the Fermi surface. In this sense Fig.~\ref{fig:Extra} should be viewed as the re-arrangement of states deep away from the Fermi surface that is responsible for the formation of the 
``core'' of the $e$ particle and this re-arrangement is additionally dressed by particle hole excitations arbitrarily close to the Fermi surface.

\begin{figure}[!t]
    \centering
    \includegraphics{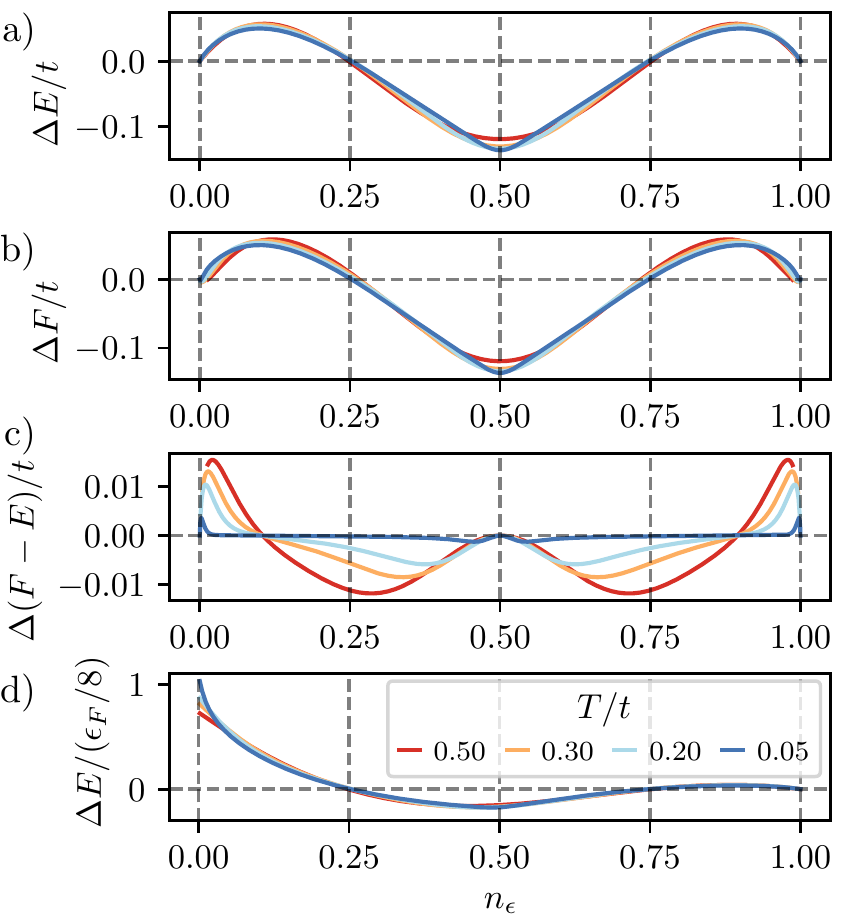}
    \caption{a) Energy difference between the $\pi$ and zero flux configurations at fixed particle number $N_{\varepsilon}$. b) Free energy difference between the $\pi$ and zero flux configurations at fixed chemical potential $\mu$. c) Difference $\Delta_{FE}$ between the energy values of a) and b). d) At fixed particle number $N_{\varepsilon}$, the lattice model recovers the result from Eq.~\eqref{eq:EnDiff-FixedN} at small fillings. Here $\epsilon_{F}=\mu-E_{0}$, with $E_{0}=-4t$ is the energy of the bottom of the band. Results for the lattice model from Eq.~\eqref{eq:FZ2-FullLatticeModel} with square geometry considering $120\times 120$ plaquettes.}
    \label{fig:Fig-12-FZ2}
\end{figure}

\subsection{Square lattice geometry}

We will now describe the behavior of the flux inserted into the Fermi liquid of $\varepsilon$ particles in the square lattice tight-binding model from Eq.~\eqref{eq:FZ2-FullLatticeModel}. To do so, we choose a square geometry with open boundaries that allows to insert a single flux, and place it at its center\footnote{It is possible to microscopically construct states with a single $e$ particle in open lattices as discussed in Ref.~\cite{RaoInti2020}, but one can also view this as an approximation when the remaining $e$ particle is taken to be very far.}. We take lattices with an odd number of plaquettes so that the flux can be placed in a unique central plaquette. Fig.~\ref{fig:Fig-12-FZ2} illustrates the energy cost to insert the flux in the thermodynamic $T\rightarrow0$ limit, which we find to remain finite regardless of the lattice filling of the $\varepsilon$ fermions. This energy cost can be equivalently determined by computing the energy difference between the states with and without flux with fixed $\varepsilon$ particle number, as shown in Fig.~\ref{fig:Fig-12-FZ2}a), or by computing the free energy difference, $\Delta F = \Delta E -\mu\Delta N$, at fixed chemical potential, as shown in Fig.~\ref{fig:Fig-12-FZ2}b). Both ways of computing provide the same energy value for the insertion of the flux in the thermodynamic limit, as shown in Fig.~\ref{fig:Fig-12-FZ2}c), which approaches zero at $T\rightarrow 0$ and support our finding that the $e$ particle remains a fully deconfined finite energy quasiparticle in the presence of a Fermi liquid of $\varepsilon$ particles. Finally, Fig.~\ref{fig:Fig-12-FZ2}d) shows that we recover the values of the disk model from Eq.~\eqref{eq:EnDiff-FixedN} at small fillings near the bottom of the band. 

\begin{figure}
    \centering
    \includegraphics*[scale=1.0]{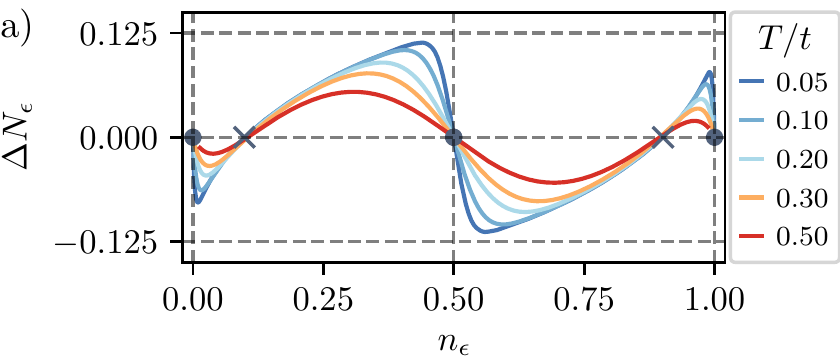}
    \includegraphics*[scale=1.0]{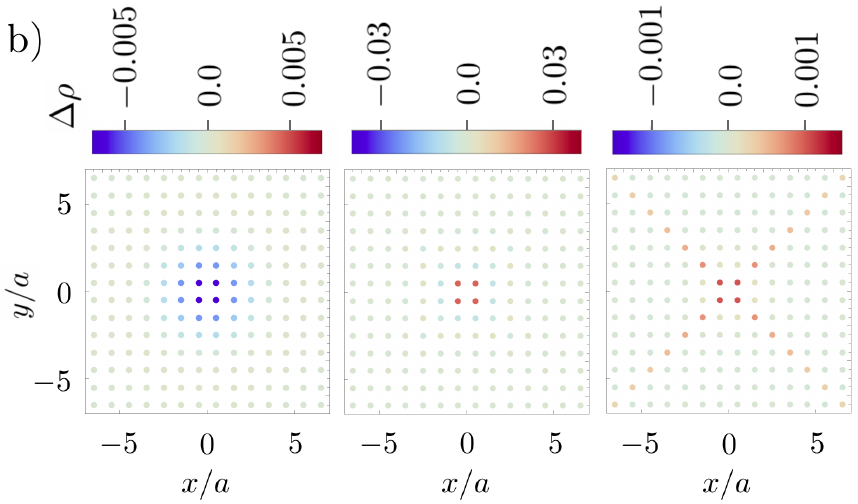}
    \caption{a) Particle number difference between the $\pi$ and zero flux configurations at fixed chemical potential $\mu$. The dots (crosses) indicate the fillings that are stable (unstable) to dilute insertions of $e$ particles. b) From left to right, local density difference between the $\pi$ and $0$-flux configurations near the fillings $n_{\varepsilon}=0,1/4$ and $1/2$, respectively. Results for the lattice model from Eq.~\eqref{eq:FZ2-FullLatticeModel} with a square geometry of $120\times 120$ with a temperature $T/t=0.05$.}
    \label{fig:Fig-13-FZ2}
\end{figure}

\begin{figure}
    \centering
    \includegraphics{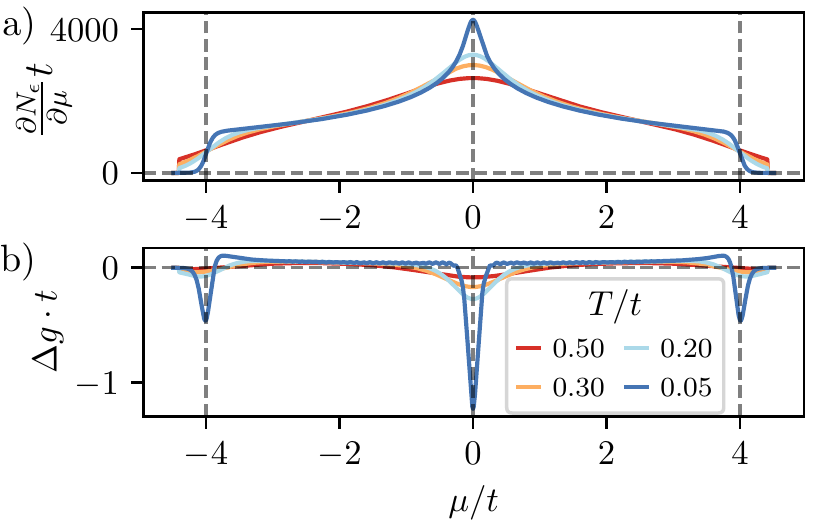}
    \caption{a) Bare thermodynamic density of states for the state with no flux, $\Phi=0$. b) Difference between the compressibility or thermodynamic density of states between the $\pi$ and zero-flux states as a function of the chemical potential of the $\varepsilon$ Fermi liquid in the square lattice. The sharp dips are expected to evolve into Dirac delta functions at zero temperature in the thermodynamic limit, sitting on top of a smooth positive background as a function of $\mu$. The delta-function dips are located at special fillings of the square lattice corresponding to $n_{\varepsilon}=\left\{ 0,1/2,1 \right\}$, at which the $\varepsilon$ Fermi liquid can accomodate a dilute amount of distant $\pi$-fluxes. Results for the lattice model from Eq.~\eqref{eq:FZ2-FullLatticeModel} with square geometry considering $120\times120$ plaquettes.}
    \label{fig:Fig-14-FZ2}
\end{figure}

Fig.~\ref{fig:Fig-13-FZ2}a) shows the particle number difference between the state with and without flux as a function of the $\varepsilon$ filling of the lattice. Generally the flux carries a screening cloud with a finite fraction of $\varepsilon$ particle fluid, but the precise value of this fraction depends on the filling of the lattice, although it approaches the value obtained for the disk, $\Delta N_{\varepsilon}=-1/8$, at small fillings near the bottom of the band. The core of this cloud is spatially localized near the flux at various fillings as shown in Fig.~\ref{fig:Fig-13-FZ2}b), although its precise shape changes with the behavior near $1/2$ filling being sharply anisotropic resembling the underlying $C_{4}$ symmetry of the lattice. Fig.~\ref{fig:Fig-14-FZ2} shows the value of the difference of the thermodynamic density of states between the two configurations with and without flux, defined as
\begin{equation}
    \Delta g = \dfrac{\partial N_{\varepsilon}(\mu,T)}{\partial\mu}\Bigg|_{\Phi=\pi} - \dfrac{\partial N_{\varepsilon}(\mu,T)}{\partial\mu}\Bigg|_{\Phi=0} \ .
\end{equation}
The thermodynamic density of states reflects the states in the energy resolved spectrum from which the screening cloud of $\varepsilon$ particles is primarily made out of. $\Delta g$ features sharp dips at fillings $0,1/2,1$ which will approach Dirac delta functions in the thermodynamic limit as shown in Fig.~\ref{fig:Fig-14-FZ2}. The magnitude of the integral under the Dirac delta functions at $0$ and $1$ tends to the value $\pm 1/8$ in the thermodynamic limit, and they confirm the picture of Fig.~\ref{fig:Extra} for the continuum parabolic model where we anticipated that the screening cloud is made primarily from states at the bottom of the band. Meanwhile, the magnitude of the integral at half-filling is $1/4$ in the thermodynamic limit. There is in addition a continuous part of the density of states difference that varies as a function of filling. The equality between energy difference at fixed particle number and free-energy difference at fixed chemical potential, which is shown in Fig.~\ref{fig:Fig-12-FZ2}c), can be understood by noting that when inserting the flux, the fraction of particles that is removed (added) by the rearrangement of the density of states deep below the Fermi level is added (removed) by occupying (emptying) states near the Fermi level when the flux is inserted at fixed particle number. This is analogous to Fig.~\ref{fig:Extra}, although in general the states are removed from various energies smoothly except near energies where the delta functions appear at the bottom, top and center spectrum of the lattice dispersion (see Fig.~\ref{fig:Fig-14-FZ2}).

\begin{figure}[!t]
    \centering
    \includegraphics{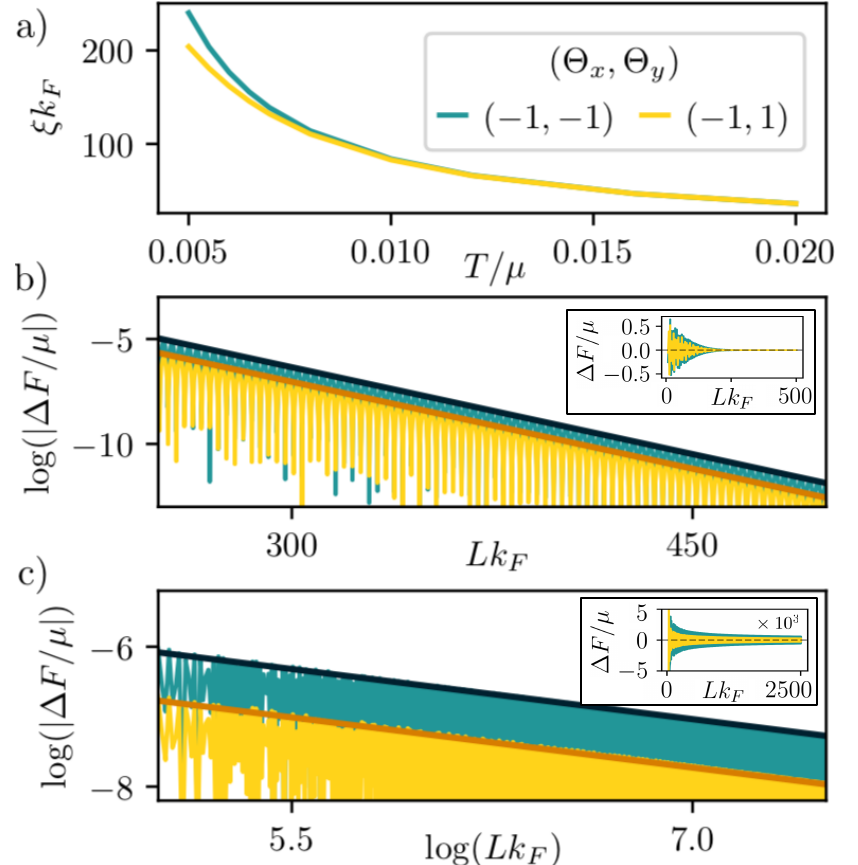}
    \caption{Free energy difference between the ground states in the torus with twisted and periodic boundary conditions. a) The system is characterized by a correlation length $\xi(T)$ that separates power-law and exponential decays. Such correlation length diverges in the zero temperature limit. b) Exponential decay with system size $L\gg\xi$ at finite temperatures, in this case for $T/\mu=0.02$. c) Power law decay of free-energy difference with system size $L$ in the zero temperature limit, in this case using $T/\mu = 6/(Lk_F)$. In b) and c), the dark blue and orange lines are linear fits to the envelope of $|\Delta F/\mu|$. These results have been obtained at fixed chemical potential for the parabolic model from Eq.~\eqref{eq:FZ2-FullLatticeModel}.}
    \label{fig:Fig-15-FZ2-TwistedBoundary}
\end{figure}

This thermodynamic density of states provides information on what are the fillings at which the $e$ particles carrying $\pi$-flux can be added to the $\varepsilon$ Fermi liquid self-consistently in a dilute fashion without changing the filling of the liquid. Specifically, for the special fillings at which $\Delta g$ is negative, the $\varepsilon$ Fermi liquid can accommodate insertions of dilute distant $e$ particles impurities in a thermodynamically stable fashion. This only occurs at the three special fillings $n_{\varepsilon}=\left\{ 0,1/2,1 \right\}$. At these fillings, marked by dots in Fig.~\ref{fig:Fig-13-FZ2}a), the $e$ particle carries a screening cloud with a zero total number of $\varepsilon$ particles, and therefore adding a dilute fraction of them does not change the $\varepsilon$ filling. Moreover, if we deviate from these precise fillings by some small amount, $n_{\varepsilon}+\delta n_{\varepsilon}$, the insertion of $e$ particles will carry an additional particle number of $\varepsilon$ particles which self-consistently drives the total filling back to the fixed value $n_{\varepsilon}$, as can be seen by the signs of $\Delta N_{\varepsilon}$. There are in addition two intermediate fillings at $0 < n_{\varepsilon} < 1/2$ (and $1/2 < n_{\varepsilon} < 1$),  marked by a cross in Fig.~\ref{fig:Fig-13-FZ2}a), where the screening cloud of $e$ particles carries zero net $\varepsilon$ particles. However, these other fillings are thermodynamically unstable to dilute insertion of $e$ particles, in the sense that for small deviations $\delta n_{\varepsilon}$ a dilute insertion of $e$ particles drives the density $n_{\varepsilon}$ away from these fillings. The stable fillings corresponding to the completely empty and filled lattice, $n_{\varepsilon}=\left\{ 0,1 \right\}$, are just the usual understood TC vacua\footnote{Strictly speaking, the $n_{\varepsilon}=1$ is a different vacuum since it leads to different projective symmetry group for translations of the $e$ particle, as defined in Ref.~[\onlinecite{wen2002quantum}], however apart from symmetry based distinctions it has the same bulk topological properties of the ordinary TC vacuum at $n_{\varepsilon}=0$}. The resulting quantum ordered states at $n_{\varepsilon}=1/2$ in the presence of $e$ particles could however lead to novel topological orders if the $e$ particles are not ordered into simple static configurations, which is an interesting future direction for further investigation.

The precise characterization of universal aspects of quantum gapless phases is largely an open problem. We would like, however, to diagnose whether there is a sense in which ground state degeneracy is preserved for the twists of boundary conditions in the case of a Fermi liquid of $\varepsilon$ particles. We do this by computing the ground state free energy difference for the Fermi liquid of $\varepsilon$ particles placed in the torus in the continuum parabolic band model at fixed chemical potential. Fig.~\ref{fig:Fig-15-FZ2-TwistedBoundary} shows that the free energy vanishes in the thermodynamic limit following a decay law of the form:
\begin{IEEEeqnarray}{rClrl}
    \Delta F/\mu & \approx & A L^{-\alpha(T)} \ & \quad \mathrm{for} & \ \ L \ll \xi(T) \ , \\
    \Delta F/\mu & \approx & B e^{-L/\xi(T)} \ & \quad \mathrm{for} & \ \ \xi(T) \ll L \ ,
\end{IEEEeqnarray}
where we assume a torus of area $L^2$. The correlation length $\xi(T)$ diverges in the zero temperature limit, as it is depicted in Fig.~\ref{fig:Fig-15-FZ2-TwistedBoundary}a). For system sizes larger than the correlation length, the free energy difference decays exponentially as it is shown in Fig.~\ref{fig:Fig-15-FZ2-TwistedBoundary}b), while for system sizes smaller than the correlation length display a power law decay as the one of Fig.~\ref{fig:Fig-15-FZ2-TwistedBoundary}c). In the $T=0$ limit, the power is $\alpha \approx 1/2$.  The energy difference at fixed particle number follows similar results. As we have reviewed in Sec.~\ref{sec:BoseZ2}, the ground state degeneracy of fully gapped topologically ordered phases comes hand-in-hand with anyon deconfinement, and what we have found is that, in a sense, this relation holds in the thermodynamic limit for the gapless Fermi liquid of $\varepsilon$ fermions.

\section{Discussion}
\label{sec:Discussion}

We have reviewed constructions that allow to rewrite the Hamiltonians of ordinary spins in terms of non-local variables in an exact manner. The idea is to exploit the Toric Code (TC) as a vacuum in which novel states can be obtained by adding to it a finite density of its anyonic quasiparticles. To do so, out of the three non-trivial TC quasiparticle types, $\left\{ e,m,\varepsilon \right\}$, one must choose two of them as the elementary building blocks and view the third as a non-elementary composite. The two building block particles have local self-statistics, namely, they are self-bosons or self-fermions, but a non-local mutual semion statistics. Therefore, if one of the two quasiparticle types is kept frozen by enforcing a local conservation law, the remaining dynamics of the particles that are allowed to fluctuate will be that of ordinary local bosonic or fermionic quantum particles. In the case when the building blocks are the two bosonic particles, $e$ and $m$, one obtains the structure of the familiar bosonic $\mathbb{Z}_2$ lattice gauge theory~\cite{fradkin2013field,wen2004quantum}. However, in the case in which one of the two building blocks is the fermionic particle, $\varepsilon$, one obtains a new kind of fermionic $\mathbb{Z}_2$ lattice gauge theory, and the construction can be viewed as a form of 2D local bosonization of fermions, as recently emphasized in Ref.~[\onlinecite{chen2018exact}]. 

We have endowed these bosonic and fermionic $\mathbb{Z}_2$ lattice gauge theories with an additional global U(1) symmetry associated to the particle number conservation of the particle that is allowed to fluctuate. In the bosonic case, this allows to construct superfluid states of these mobile particles. Then, the anyon that remains immobile acts as a $\pi$-flux that binds a half-vortex in the superfluid, as pointed out in Ref.~[\onlinecite{kivelson1989statistics}]. Such $\pi$-flux remains only marginally confined in a logarithmic fashion, in contrast to the strong linear confinement that occurs in the confined phase of ordinary bosonic $\mathbb{Z}_2$ lattice gauge theory, which is implicitly assumed in formal anyon ``condensation'' schemes~\cite{diamantini1996gauge,hansson2004superconductors} without a global U(1) symmetry. 

In the fermionic case, the additional global U(1) symmetry for the $\varepsilon$ particles allows to construct naturally Fermi liquid states. These states share many universal properties with orthogonal metals~\cite{nandkishore2012orthogonal}, but with the distinction that in our case the U(1) symmetry is not the microscopic electron number conservation. Remarkably, we have found that in these states the immobile $\pi$-fluxes have a finite energy cost even though they induce a long-ranged power-law-decaying oscillatory disturbance of the Fermi fluid, akin to Friedel oscillations. Therefore, the $\pi$-fluxes are finite energy fully deconfined quasiparticles.

We have studied the detailed properties of $\pi$-fluxes embedded in the fermion liquid of $\varepsilon$ particles, and have shown that they are accompanied by a finite screening cloud of $\varepsilon$ fermions. In the case of parabolic bands, this cloud contains exactly $1/8$ of a fermionic hole. In general, the precise number of particles making up this cloud changes as a function of details of the lattice dispersion and fillings of the lattice, and therefore is not expected to be universal. However, these distinctive characteristics surrounding the $\pi$-flux could be useful in local spectroscopic studies searching for orthogonal metals, where the screening cloud would manifest as a characteristic spatial profile of the electronic charge distribution near the $\pi$-flux.

\section{Acknowledgements}
We are thankful to Vijay Shenoy for valuable discussions. O.P. is thankful to M. A. H. Vozmediano and F. de Juan for support and useful conversations. O.P. is supported by an FPU predoctoral contract from MECD No. FPU16/05460 and the Spanish grant PGC2018-099199-BI00 from MCIU/AEI/FEDER.

\bibliography{Manuscript}

\end{document}